\documentclass[nonacm]{acmart}

\AtBeginDocument{%
  \providecommand\BibTeX{{%
    \normalfont B\kern-0.5em{\scshape i\kern-0.25em b}\kern-0.8em\TeX}}}

\usepackage[english]{babel} 

\usepackage{url,hyperref,lineno,microtype,tabularx} 
\usepackage{xcolor}
\usepackage{multirow}
\usepackage{multicol}
\usepackage[normalem]{ulem}
\usepackage{booktabs}

\newif\ifworkinprogress
\workinprogressfalse

\ifworkinprogress
	\newcommand{\ms}[1]{\textcolor{blue}{\textbf{[MS:] #1}}}
	\newcommand{\chb}[1]{\textcolor{magenta}{\textbf{[ChB:] #1}}}
	\newcommand{\el}[1]{\textcolor{violet}{\textbf{[EL:] #1}}}
	\newcommand{\dk}[1]{\textcolor{ForestGreen}{\textbf{[DK:] #1}}}
	\newcommand{\rei}[1]{\textcolor{RawSienna}{\textbf{[WR:] #1}}}

\else
  \newcommand{\ms}[1]{}
  \newcommand{\chb}[1]{}
  \newcommand{\el}[1]{}
  \newcommand{\dk}[1]{}
   \newcommand{\rei}[1]{}
\fi

	\newcommand{\changed}[1]{\textcolor{black}{#1}}
	\newcommand{\changedRev}[1]{\textcolor{black}{#1}}

\setcopyright{none}

\author{Markus Schedl}
\authornote{Corresponding author: Markus Schedl, \href{mailto:markus.schedl@jku.at}{markus.schedl@jku.at}}
\email{markus.schedl@jku.at}
\affiliation{%
  \institution{Johannes Kepler University Linz, Institute of Computational Perception, Multimedia Mining and Search Group}
  \institution{Linz Institute of Technology, AI Lab, Human-centered AI Group}
  \city{Linz}
  \country{Austria}
}
\author{Christine Bauer}
\email{c.bauer@uu.nl}
\affiliation{%
  \institution{Utrecht University}
  \city{Utrecht}
  \country{The Netherlands}
}
\author{Wolfgang Reisinger}
\affiliation{%
  \institution{Johannes Kepler University Linz, Institute of Computational Perception, Multimedia Mining and Search Group}
  \city{Linz}
  \country{Austria}
}
\author{Dominik Kowald}
\affiliation{%
  \institution{Graz University of Technology and Know-Center}
  \city{Graz}
  \country{Austria}
}
\author{Elisabeth Lex}
\affiliation{%
  \institution{Graz University of Technology and Know-Center}
  \city{Graz}
  \country{Austria}
}

\renewcommand{\shortauthors}{[Preprint version as of \today]}

\begin{abstract}
\noindent
Music preferences are strongly shaped by the cultural and socio-economic background of the listener, which is reflected, to a considerable extent, in country-specific music listening profiles. 
Previous work 
has already identified several 
 country-specific differences in the popularity distribution of music artists listened to. 
In particular, what constitutes the ``music mainstream'' strongly varies between countries.
To complement and extend these results, the article at hand delivers the following major contributions: 
First, using state-of-the-art unsupervised learning techniques, 
we identify and thoroughly investigate (1)~country profiles of music preferences on the fine-grained level of music tracks (in contrast to earlier work that relied on music preferences on the artist level) 
and (2)~country archetypes that subsume countries sharing similar 
patterns of listening preferences.
Second,~we formulate four user models that leverage the user's country information on music preferences.
Among others, we propose a user modeling approach to describe a music listener as a vector of similarities over the identified country clusters or archetypes.
Third,~we propose a context-aware 
music recommendation system that leverages implicit user feedback, where context is defined via the four user models.
More precisely, it is a multi-layer generative model based on a variational autoencoder, in which contextual features can influence recommendations through a gating mechanism.
Fourth,~we thoroughly evaluate the proposed recommendation system and user models on a real-world corpus of more than one billion listening records of users around the world (out of which we use 369 million in our experiments) and show its merits vis-\`a-vis state-of-the-art algorithms that do not exploit this type of context information.
\end{abstract}

\begin{document}

\title{[Preprint version] Listener Modeling and Context-aware Music Recommendation Based on Country Archetypes} 

\maketitle


 \section{Keywords:} 
  music, recommender system, culture, country, clustering, context, user modeling, music preferences.

\section{Introduction}\label{sec:intro}
Recommendation systems (or recommender systems) have become 
an important means to help users find and discover 
various types of content and goods, including movies, videos, books, and 
food~\cite{ricci_etal:rsh:2015}. 
As such, they represent substantial business value.
In the music industry, recommender systems---powered by machine learning and artificial intelligence---have radically changed the market; they have even become major drivers in this industry. Essentially, music recommender systems (MRS) shape today's digital music distribution~\cite{schedl_etal:rsh:2015} and have become vital tools for marketing music to a targeted audience, as evidenced by the success of recommender-systems-featuring music streaming services such as Spotify, Deezer, or Apple Music. 
While MRS operate in a multi-stakeholder environment including platform providers, artists, record companies, and music consumers/listeners~\changed{\cite{bauer_impactrs2019}}, 
it is most commonly the music consumers/listeners, who are considered the users of an MRS. In the paper at hand, we also take this perspective.

Traditionally, content-based filtering (CBF) and collaborative filtering (CF)---or hybrid combinations thereof---have been the most common algorithms to create recommender systems \cite{ricci_etal:rsh:2015}. The former assumes that users will like items similar to the ones they liked in the past, and therefore selects items to recommend according to some notion or metric of similarity in terms of item content ({e.g.}, music style, timbre, or rhythm) between the user's liked items and unseen items from the catalog. 
In contrast, CF assumes that a user will prefer items that are liked by other users with similar preferences. In this case, items to recommend are, for instance, found by comparing the target user's consumption or rating profile to that of the other users, identifying the most similar other users, and recommending what they liked (user-based CF).
Alternatively, users and items can be directly matched via similarities computed in a joint low-dimensional representation of users and items ({i.e.}, model-based CF).

Enhancing the classical approaches CF and CBF, in recent years, researchers started to leverage additional information---beyond users, items, and their interactions---to improve recommendations. Recommender systems that consider user characteristics or information describing a situation are typically referred to as \emph{context-aware recommendation systems}~\cite{adomavicius_tuzhilin:rsh:2015}. Next to considering time and location as contextual side information, taking information derived from the user's country into account has been demonstrated to improve recommendation quality; for instance, cultural and socio-economic characteristics of the user's country~\cite{zangerle_etal:umap:2018}, or the user taste's proximity to their country-specific music mainstream (``mainstreaminess'')~\cite{10.1371/journal.pone.0217389}.

Against this background, we approach the task of context-aware music recommendation based on country information; in contrast to most previous works, 
we consider user country in our approach without using any external information about the country, such as cultural, economic, or societal information. The reason is that respective data sources about countries (e.g., Hofstede's cultural dimensions,\footnote{\url{https://geerthofstede.com/research-and-vsm/dimension-data-matrix}} the Quality of Government measures,\footnote{\url{https://qog.pol.gu.se}} or the World Happiness Report\footnote{\url{https://worldhappiness.report}}) 
provide information on the country level, which may not necessarily reflect the circumstances of individual users and, thus, 
can introduce problems in the recommendation process. For instance, 
cultural values or income may be very unequally distributed among a country's population. 

To avoid this, instead of using external information derived from the user's country, we leverage purely the self-reported country information of the users as available in the system, and investigate how behavioral data about music listening can be used to (1)~identify archetypal country clusters based on track listening preferences, (2)~how users can be modeled using the results of~(1), and (3)~how the resulting user models can be integrated into a state-of-the-art deep learning-based music recommendation algorithm. 

As in many other domains, nowadays, deep neural network architectures dominate research in music recommendation systems, due to their ability to automatically learn features from low-level audio signals and their superior performance~\cite{10.3389/fams.2019.00044}. This article is no exception. We propose a multi-layer generative model in which contextual features can influence recommendations through a gating mechanism. 

In this context, we formulate the following research questions:

\textbf{RQ1:} To what extent can we identify and interpret groups of countries that constitute \textit{music preference archetypes}, from behavioral traces of users' music listening records?

\textbf{RQ2}: Which are effective ways to model the users' geographic background as a contextual factor for music recommendation? 

\textbf{RQ3}: How can we extend a state-of-the-art recommendation algorithm, based on variational autoencoders, to include user context information, in particular, the geo-aware user models developed to answer RQ2? 

In the remainder of this article, we first explain the conceptual foundation of our work and discuss it in the context of related research (Section~\ref{sec:relatedwork}).
Subsequently, we detail the methods we adopt to investigate the research questions; in particular, we specify the approaches used for data preparation, clustering, user modeling, and track recommendation (Section~\ref{sec:methods}).
The results of our experiments on uncovering geographic music listening archetypes and on music track recommendation, altogether with a detailed discussion thereof, are presented in Section~\ref{sec:results}. 
Finally, Section~\ref{sec:conclusions} concludes the article with a brief summary of the major findings, a discussion of limitations, and pointers to future work.

\section{Conceptual Background and Related Work}\label{sec:relatedwork}

\changed{A multitude of factors have been found to influence an individual's music preferences. 
There is a long history of research investigating the relationships between music preferences and, for instance, demographics~\cite{BonnevilleRoussy2013,colley2008,Cheng:2017:EUI:3077136.3080772}, personality traits~\cite{Rentfrow_doremi,schafer2017can}, and social influences~\cite{terbogt2011,BonnevilleRoussy2018_socialinfluences} }. 



\changedRev{In the middle of the nineteenth century emerged a cultural hierarchy in America~\cite{Dimaggio1982,levine1988highbrow} 
where a high social status patronized the fine arts (referred to as ``highbrow'') while all other forms of popular culture were associated with a lower status (referred to as "middlebrow" or ``lowbrow''). 
In the 1990s, a series of studies~\cite{peterson1992seven,peterson_kern_1996} have \changedRev{defended the view} that, for the elite, highbrow was being replaced by a consumption pattern termed ``omnivorousness''. Cultural omnivorousness reflects that people's taste includes both elite and popular genres.
This was subsequently shown to hold for various countries \cite[e.g.][]{Coulangeon2003,FISHER200369,Holbrook2002}. 
Also, the consumption practices of low status taste were reconceptualized: The earlier view that the lowbrow group would be willing to consume any entertainment on offer~\cite{horkheimer1972dialectic} was replaced by the finding that low status people tend to choose one form of entertainment and avoid others~\cite{bryson1997univores}. Thus, overall the view evolved from highbrow--lowbrow to omnivore--univore. 
Analyzing music consumption across eight European countries, \citet{coulangeon:hal-01053502} supported the ``omnivore--univore'' scheme rather than the former ``highbrow--lowbrow'' model.
The omnivorous cultural taste was later found unstable over time~\cite{ROSSMAN2015139}, though.
\citet{Katz-Gerro2002} has shown that the dividing line of class distinctions varies across countries and also the genre associations to social classes deviate. She concludes that, while class matters, the main determinants of cultural preferences relate to gender, education, and age \cite{Katz-Gerro1999}. 
\citet{Coulangeon2005} questions the earlier view on the reasons for the different tastes of higher- and lower-status classes: He challenges that it would be the upper class' familiarity with the so-called ``legitimate'' culture and the little accessibility to that culture for the lower-status classes, that distinguished what the upper class and lower-status classes prefer. Instead, he attributes it to the diversity of the stated preferences of people of the upper class, whereas the preferences of members of lower-status classes appear more exclusive. Later work, studying music taste in the ``modern age''~\cite{Nuccio2018}, found little evidence that musical taste is indeed aligned with class position.}


\changed{
Although there is a multitude of factors that influence an individual's music preferences that lead to a diversity of music created and listened to, there are (market) structures and other mechanisms that effect certain tendencies in what music is preferred within a particular community.}
\changed{For instance, }
the music recording industry is typically considered a globally oriented market~\cite{dolata2013transformative}\changed{
. Yet,} studies have revealed the existence of national boundaries~\cite{bauer_schedl_hicss2018}. There are various country-specific mechanisms that affect an individual's music preferences and consumption behavior: Preferences are culturally shaped~\cite{baek2015,Budzinski2017}; music perceptions vary across cultures, for instance, with respect to mood~\cite{jinhalee2014,Morrison2009,Singhi2014_ismir,stevens2012}; and countries have substantially different national market structures with respect to, for instance, available music repertoire due to copyright and licensing, advertising campaigns, local radio airplay, or quotas for national artists~\cite{hracs2017_musicproduction_consumpation,gomez2018_copyright}.

Knowledge about country-specific differences in music preferences can be explicitly used to improve music recommender systems, for instance, by leveraging information about the users' geographic or cultural background.
For instance, \citet{Vigliensoni2016} use a factorization machine approach for matrix factorization and singular value decomposition to integrate---amongst others---a user's country as context information. 
\citet{10.1371/journal.pone.0217389} use a contextual pre-filtering approach~\cite{adomavicius_tuzhilin:rsh:2015}, where the user base is first segmented by user country, and a target person is then compared to other people from the very same country (in contrast to a comparison with the entire user base). \citet{SANCHEZMORENO2016234} use a k-nearest neighbor (k-NN) approach integrating, amongst others, the user's country as attribute. \citet{zangerle_etal:umap:2018} leverage further country-specific data sources; for each country, they use the respective scores on the cultural dimensions by Hofstede~\cite{hofstede2005cultures} as well as the scores of the World Happiness Report~\cite{helliwell2016world} to tailor recommendations to the individual.

The work at hand differentiates from related work in several aspects.
\begin{itemize}
    \item First, although music preferences vary across countries, several studies \cite[e.g.][]{10.1371/journal.pone.0217389,pichl2017_ism,schedl2017_ism,moore_etal:ismir:2014} have shown similarities in music preferences between countries, typically identified with clustering approaches. 
    Yet, to the best of our knowledge, the work at hand is the first one to integrate information on country similarities into the music recommendation approach. 
    \item Second, while other work, most notably, \citet{zangerle_etal:umap:2018}, 
    reaches out to include external data about countries (such as economic factors, happiness index, cultural dimensions), the approach at hand remains independent from any external data sources, enabling platform providers to build a self-sustaining recommendation system. Such a system 
    can rely exclusively on data that is contained in the provider's platform, including users' self-disclosed country information.
    \item Third, most existing research on music preferences and recommender systems considers music preferences on a genre level \cite[e.g.][]{Skowron2017_genreprediction,ADIYANSJAH201999} or artist level \cite[e.g.][]{10.1371/journal.pone.0217389,SANCHEZMORENO2016234}. 
    Research on country-aware music recommendation systems that provide recommendations on the track level is rare \cite[e.g.][]{zangerle_etal:umap:2018}.
    However, the genre and the artist level may be too coarse-grained to reflect users' music preferences, for several reasons. Music genres are vaguely defined \changed{\cite{Beer2013,SONNETT201638,VLEGELS201776}} and users' perceptions thereof differ tremendously \changed{\cite{Brisson2019_genre,VANVENROOIJ2009315}}. Artists frequently cover several music styles throughout their career, where some tracks may be more favored than others for reasons including 
    lyrics quality, 
    the influence of associated music videos, 
    over-exposure, 
    or associations with unpleasant personal experiences \cite{cunningham_etal:ismir:2005}. 
    Accordingly, the work at hand investigates music recommendations on the track level to 
    reflect users' preferences in a \changed{more} fine-grained manner \changed{than genre labels attributed to an artist's overall repertoire could do}.
    
    \item Fourth, while deep learning approaches are increasingly used for recommender systems in general and for music recommendation in particular, the integration of geographic aspects---\changed{especially user country}---
    with deep learning for music recommendation is a particular asset of the work at hand. For instance, a recent survey on deep learning-based recommender systems \cite{Batmaz2019_ai_review} reports that extant research mainly uses textual information to capture context in approaches to context-aware recommender systems. 
    The authors particularly consider context that is extracted from items (e.g., text documents) instead of users.
\end{itemize}

\section{Methods}\label{sec:methods}
In the following, we detail how we gather and process the dataset used in our study, which contains information about users' music listening behavior (Section~\ref{sec:data}).
We then describe our approach to identify country clusters based on this dataset (Section~\ref{sec:clusters}). Finally, we elaborate on our approaches to create user models incorporating country information and we detail our neural network architecture that integrates these models (Section~\ref{sec:um_rs}).

\subsection{Data Acquisition and Processing}\label{sec:data}
\begin{figure}[h]
\begin{center}
\includegraphics[width=1\linewidth]{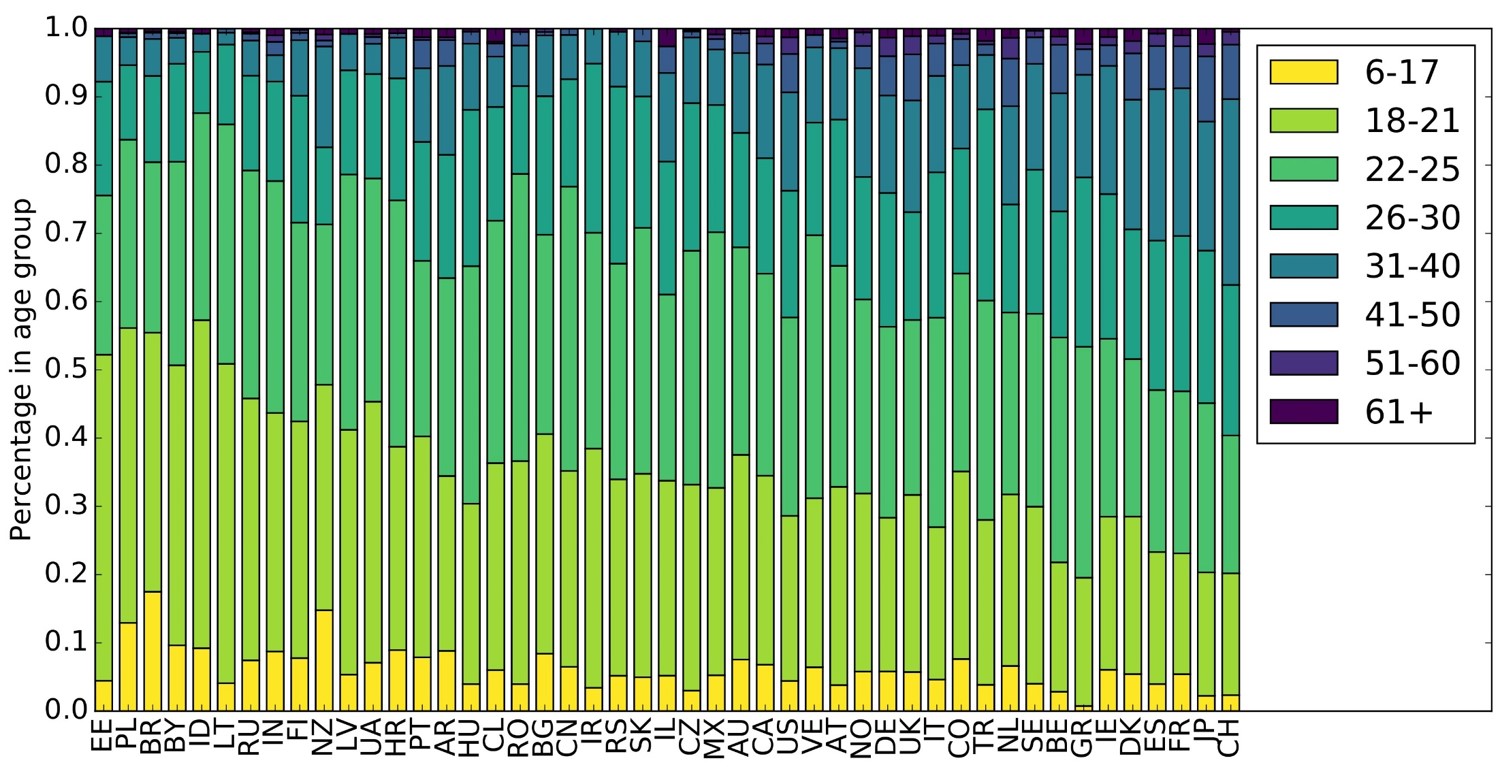}%
\end{center}
\caption{\changedRev{Distribution of age over countries. Countries are sorted in decreasing order of number of users from left to right.}\label{fig:age_vs_countries}}
\end{figure}

\begin{figure}[h]
\begin{center}
\includegraphics[width=1\linewidth]{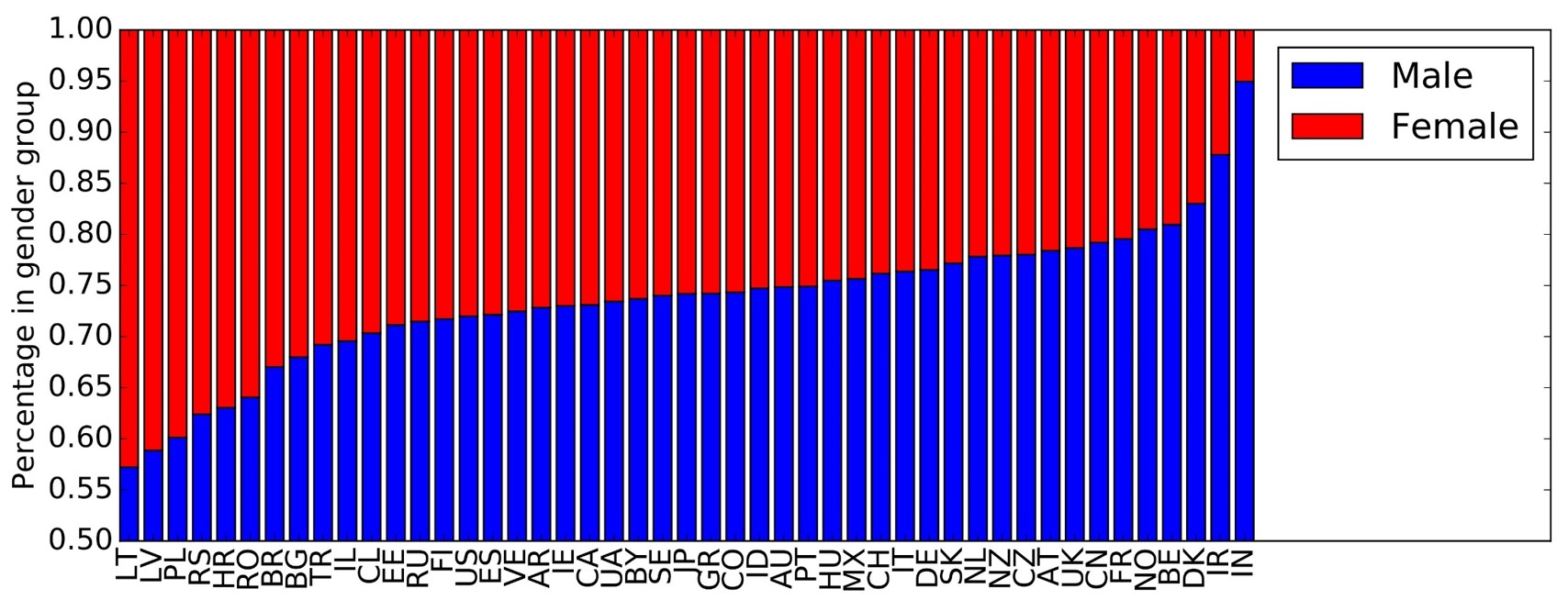}%
\end{center}
\caption{\changedRev{Distribution of gender over countries. Countries are sorted in decreasing order of number of users from left to right.}\label{fig:gender_vs_countries}}
\end{figure}

\changedRev{We base our investigations on the LFM-1b dataset \cite{Schedl:2016:LDM:2911996.2912004}, which we filter according to our requirements as detailed below.
The LFM-1b dataset\footnote{\url{http://www.cp.jku.at/datasets/LFM-1b}} contains music listening information for 120,322 Last.fm users, totaling to 1,088,161,692 individual listening events (LEs) generated between January 2005 and August 2014\changed{; the majority of LEs was created during years 2012--2014}.\footnote{The LFM-1b dataset used in our study is considered derivative
work according to paragraph 4.1 of Last.fm's API Terms of Service (\url{https://www.last.fm/api/tos}). The Last.fm Terms of Service further grant us a license to use this data (according to paragraph 4).}
Each LE is characterized as a quintuple of user-id, artist-id, album-id, track-id, and timestamp. 
The average number of LEs per user in the dataset is 8,879 (std.~15,962).
For some users, also demographic data (country, age, and gender) is available in LFM-1b.
More precisely, 46\% of the users do provide information about their country, the same percentage do provide information about their gender, and 62\% about their age.
The majority of users who provide their country are from the US (18.5\%), followed by Russia (9.1\%), Germany (8.3\%), the UK (8.3\%), Poland (8.0\%), Brazil (7.0\%), and Finland (2.6\%).
The mean age of the users who reveal it is 25.4 years (std.~9.4); the median age is 23~years.
The age distribution differs significantly between countries, though. In Figure~\ref{fig:age_vs_countries}, we show the age distribution for the countries with at least 100 users (47 countries), categorized into age groups. The youngest users are found in Estonia and Poland, while the oldest users are Swiss and Japanese.
Among the users who indicate their gender, 72\% are male and 28\% are female. These percentages differ, however, considerably between countries. In Figure~\ref{fig:gender_vs_countries}, we therefore depict the ratios between genders, again for the top 47 countries in terms of number of users. While the Baltic countries Lithuania and Latvia have an almost equal share of male and female users, India and Iran show a very unequal distribution (around 90\% male users).
}

\begin{figure}[h]
\begin{center}
\includegraphics[width=.7\linewidth]{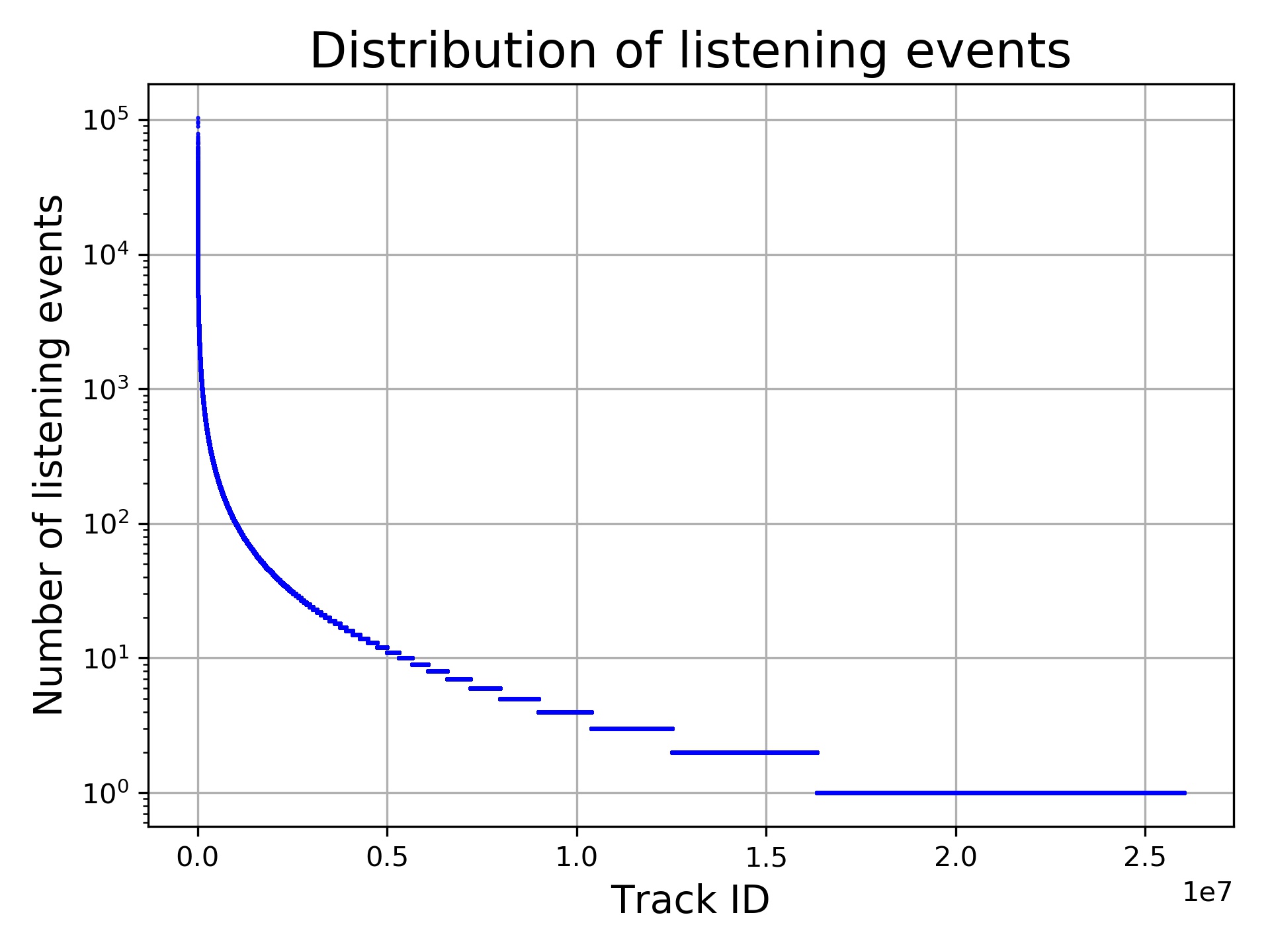}%
\end{center}
\caption{Distribution of number of listening events over all tracks (semi-log-scaled). Track identifiers are ordered by number of LEs.\label{fig:le_track_distribution}}
\end{figure}

As reported above, about 46\% of users in the LFM-1b dataset disclose their country. For our country-specific analysis, we therefore only consider users (and their LEs) for whom country information is available. 
This results in a dataset of 55,186 users, who have listened to a total of 26,021,362 unique tracks. 
The distribution of the number of LEs over tracks is visualized in Figure~\ref{fig:le_track_distribution}.

\begin{figure}[h]
\begin{center}
\includegraphics[width=.7\linewidth]{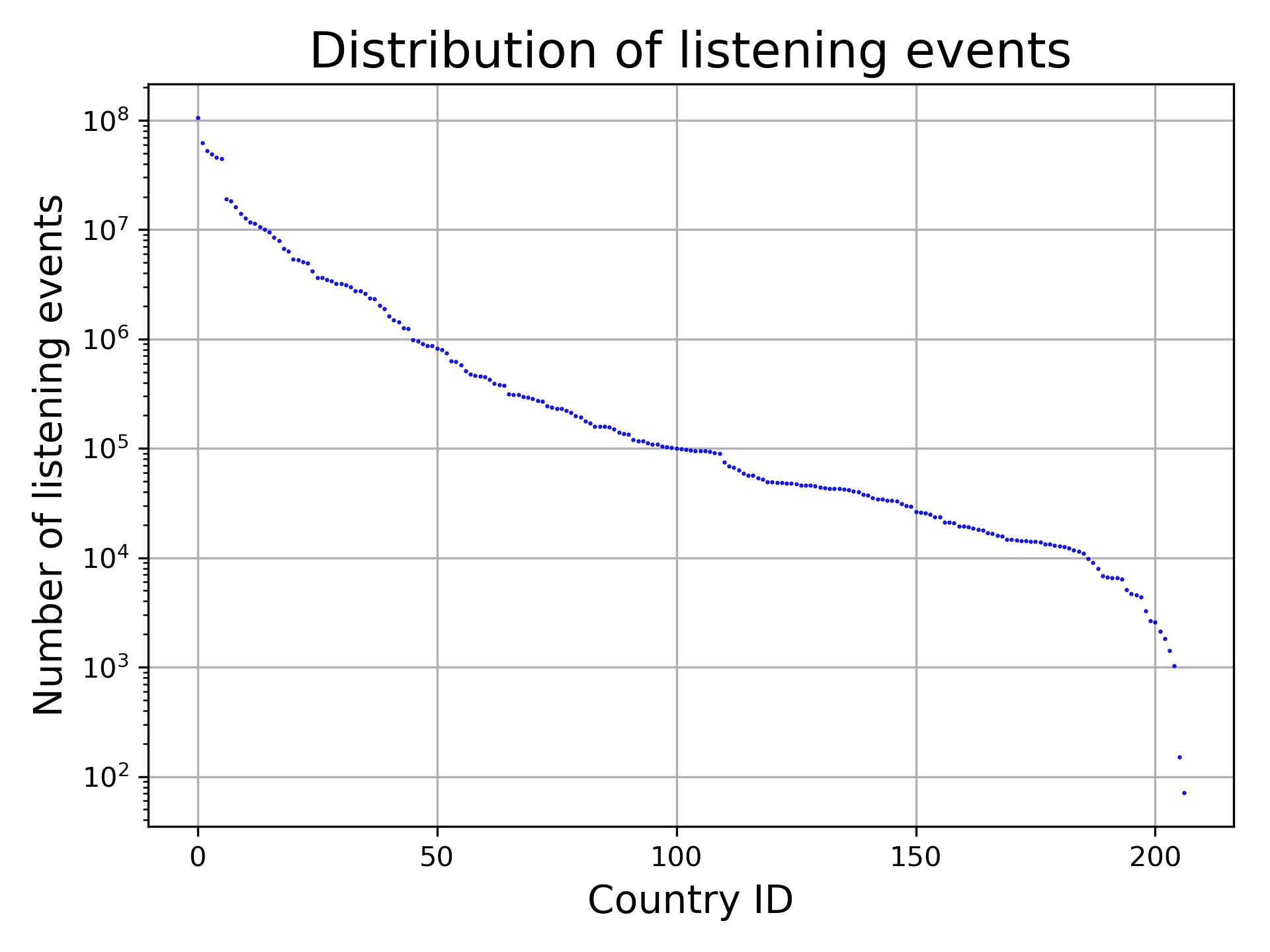}%
\end{center}
\caption{Distribution of number of listening events over all countries (semi-log-scaled). Country identifiers are ordered by number of LEs.\label{fig:les_country_distribution}}
\end{figure}

\begin{figure}[h]
\begin{center}
\includegraphics[width=.7\linewidth]{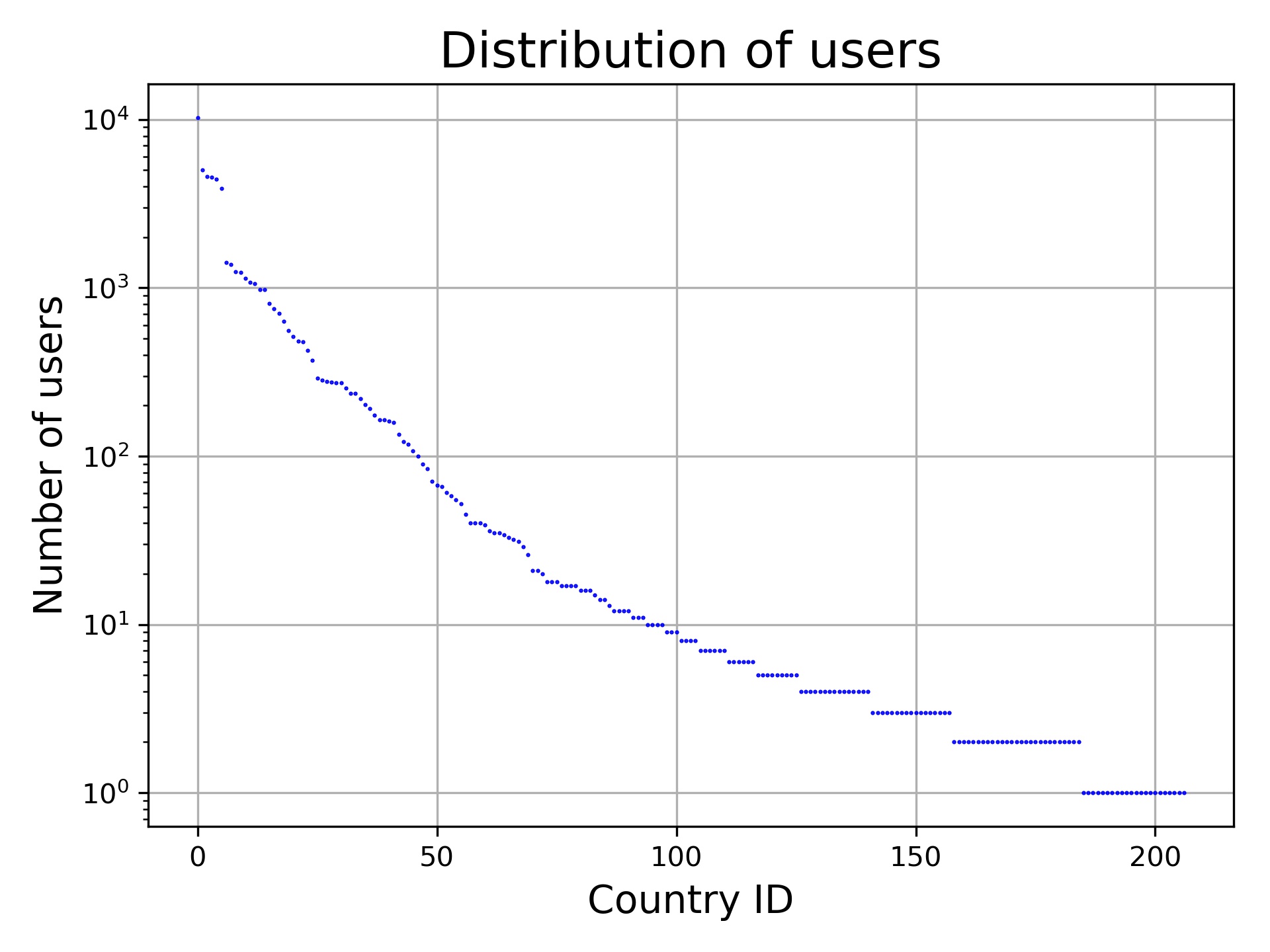}%
\end{center}
\caption{Distribution of number of users over all countries (semi-log-scaled). Country id ordered by number of users.\label{fig:users_country_distribution}}
\end{figure}

We subsequently reduce the data \changed{to decrease noise originating from the user-generated nature of the metadata in the LFM-1b dataset (in particular, misspellings and ambiguities)}, i.e., we filter out tracks and countries.
\changed{This noise would otherwise likely cause distortions in future steps of our approach.}
First, we drop tracks that have been listened to less than 1,000 times, globally, resulting in a total of 122,442 tracks to consider further.
Second, to minimize possible distortions caused by countries with a low number of LEs or a low number of unique users, we only consider countries with at least 80,000 LEs and at least 25 users.
We chose these values as thresholds based on an empirical investigation 
of the distributions of LEs and of users over countries (cf.~Figures~\ref{fig:les_country_distribution} and~\ref{fig:users_country_distribution}, respectively). The former shows a flat characteristic around country-id 100, followed by a clear gap between country-id 110 and 111 (which corresponds to 80,000 LEs). The latter reveals a sudden drop at country-id 70 (which corresponds to 25 users).
Applying this country filtering eventually results in 70 unique countries and a total of 369,290,491 LEs, which represents only a small drop of 1.5\% (in comparison to 374,770,382 LEs created by users of all countries in the dataset).
After these preprocessing steps, each country is represented as a 122,442-dimensional feature vector containing the LEs over all tracks.

%
%

\subsection{Identifying Country Clusters and Archetypes}\label{sec:clusters}
To cluster countries according to their citizens' listening behavior, it is important to first normalize the data of each country to avoid distortions caused by different country sizes. To this end, \changed{we normalize each country's feature vector to sum up to one.\footnote{\changed{Please note that country-specific results may still be influenced by some users showing particularly high playcounts. Nevertheless, we decided against excluding or penalizing the listening information of such users just because users with a high playcount indicate a more pronounced inclination to listen to music. Our reasoning is that users who contribute only few listening events to Last.fm should be considered less important to model their country-specific listening behavior than users who heavily contribute. \changedRev{In addition, removing such ``power listeners'' would distort the original distribution of users’ playcounts in the sample.}}}}
We next apply truncated SVD/PCA~\cite{halko2011finding}, reducing the dimensionality of the feature vectors to 100, while still preserving 99.8\% of the variance in the data.\footnote{Reducing the dimensionality of the dataset to 50 dimensions preserves only 90.1\% of the variance.}
Taking these 100-dimensional feature vectors as an input to a t-distributed Stochastic Neighbor Embedding (t-SNE)~\cite{vandermaaten:jmlr:2008} and subsequently using OPTICS~\cite{Ankerst:1999:OOP:304182.304187} enables us to visualize the data and identify clusters of countries sharing similar music listening behaviors.

\changed{
T-SNE is a visualization technique that embeds high-dimensional data in a low-dimensional (typically, two-dimensional) visualization space, paying particular attention to preserving the local structure of the original data. It is particularly useful to disentangle data points that lie on more than one manifold.
T-SNE represents proximities or affinities between pairs of data items by estimating the probability that the first data item will choose the second one as its nearest neighbor, and vice versa. 
In the original data space, this probability is modeled by means of a Gaussian distribution centered around each data item in the high-dimensional space; in the visualization space by means of a t-student distribution centered around each data item in the low-dimensional space.
Kullback-Leibler divergence of the joint distributions between pairs of data points in the original space and in the visualization space is then minimized via gradient descent.}

\changed{
OPTICS (Ordering Points To Identify the Clustering Structure) is a density-based clustering method that creates a linear ordering of data items based on their \changedRev{spatial} proximity.
For this purpose, OPTICS first identifies core data points that have at least a certain number of neighbors in their vicinity (the minimum cluster size) and assigns a core distance to them, describing how dense the area around each core point is. 
Furthermore, a reachability distance between each pair of data items $o$ and $p$ is established, which is the maximum of (1) the distance between $o$ and $p$ and (2) the core distance of $o$, whichever is bigger. Data items assigned to the same cluster have a lower reachability distance to their nearest neighbors than items that belong to different clusters.
OPTICS subsequently creates an ordering of data items in terms of their reachability distance and identifies sudden changes in reachability between neighboring items, assuming that these correspond to cluster borders. The number of clusters is controlled by a parameter $\xi$ that defines the minimum steepness (relative change in distance) between neighboring data items to be considered a cluster boundary.\footnote{\changed{In this work, we use Euclidean distance as distance metric and set $\xi=0.05$.}}
}

\changed{As for parameter optimization,} we adopt a grid search strategy to identify a well-suited perplexity for t-SNE~(5) and a minimum size of clusters, \changed{i.e.,~minimum number of data items in each cluster,} for OPTICS~(3).\footnote{\changed{More precisely, we performed 
grid search on t-SNE perplexity in the range [1, 2, 3, 5, 10, 15, 20, 25, 30, 35, 40, 50] and on the minimum number of data points per cluster enforced by OPTICS in the range [2, 3, 4, 5], optimizing for average neighborhood preservation ratio (nearest neighbor consistency).}}
Please note that we use ISO 3166 2-digit country codes to refer to countries in this article\footnote{\url{https://www.iso.org/iso-3166-country-codes.html}}.

\begin{figure}[h]
\begin{center}
\includegraphics[width=.7\linewidth,trim=0cm 0cm 0cm 0cm,clip]{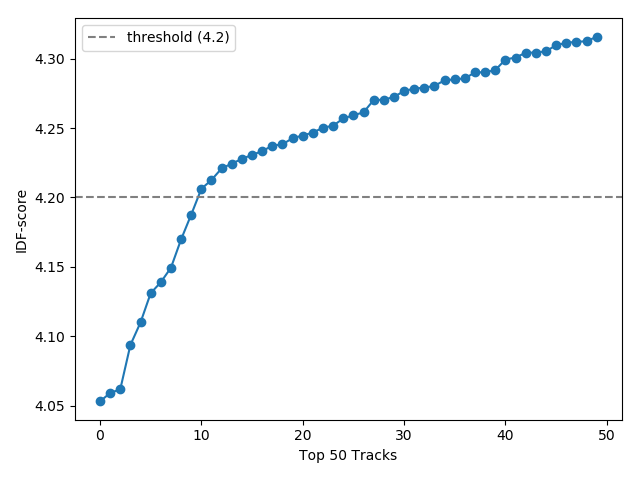}
\end{center}
\caption{\label{fig:idf_dist}Inverse document frequency (IDF) scores for the top 50 tracks.}
\end{figure}

For an analysis of the identified clusters in a way that enables the establishment of archetypes of music preferences, we adopt the following approach. As shown in Figure~\ref{fig:le_track_distribution}, we observe a long-tail distribution of listening events over tracks, which means that a few dominating tracks are listened to by a lot of users, while most tracks are only listened to by a few users. Thus, these dominating tracks will also be popular among the list of top-tracks per cluster, which makes it hard to distinguish between the clusters and to interpret their corresponding archetypes. To overcome this, we adapt a scoring function similar to the inverse document frequency (IDF)~\cite{Jones72astatistical} metric from the field of information retrieval, which assigns high scores to rarely occurring tracks and low scores to frequently occurring tracks. Formally, we define IDF for each track $t_i$ as $IDF(t_i) = \log_{10} \frac{N}{n_i}$, where $N$ is the number of all listening events and $n_i$ is the number of LEs for track $t_i$.
The distribution of IDF values of the top 50 tracks, in terms of $IDF(t_i)$, is plotted in Figure~\ref{fig:idf_dist}. 
In an empirical analysis, we identify 10 overall dominating tracks using a threshold of 4.2 on the IDF values (see Figure~\ref{fig:idf_dist}). These tracks are Rolling in the Deep by Adele, Somebody That I Used to Know by Gotye, Islands and Intro by The xx, Blue Jeans by Lana Del Rey, Supermassive Black Hole by Muse, Skinny Love by Bon Iver as well as Use Somebody, Sex on Fire and Close by Kings of Leon. We remove these tracks from further analyses when discussing archetypes as these 
are not suited to discriminate between clusters. 

In our analysis of archetypes, we include genre annotations, which we obtain as follows. 
For all tracks in the dataset, we retrieve the top user-generated tags using the Last.fm API.\footnote{\url{https://www.last.fm/api/show/track.getTopTags}} Subsequently, we filter the tags of each track using a comprehensive list of music genres and styles from Spotify, called Spotify microgenres~\cite{spotify_microgenres2018}.
This list contains 3,034 genre names \changed{(as of May 2019 when we extracted them)}, 
including umbrella genres 
\changed{such as} pop and country, as well as smaller niches such \changed{as} Thai hip-hop, German metal, and discofox~\cite{spotify_microgenres2018}. The fine-grained reflection of subtle differences in microgenres \changed{provides a more particularized basis for describing the clusters, compared to the use of a more coarse-grained taxonomy of music genres. We note as a limitation that the microgenre categories are defined in a similarly vague manner as coarse-grained taxonomies of music genre \cite{Beer2013,SONNETT201638,VLEGELS201776}; and the semantics associated with (micro)genre names have evolved over time so that a precise definition appears difficult. Relying on a big corpus of data where microgenres are visualized and sonified (see The Every Noise project\footnote{\url{http://everynoise.com}}), we nevertheless believe that using the concept of microgenres helps future research to build upon our work. Further note that we rely on the top user-generated tags from the Last.fm community for attributing microgenres to tracks; the microgenre--track associations, thus, reflect the Last.fm community's understanding of microgenres, which may not be congruent with the music experts' understanding.} 
\changedRev{Additionally, synonyms may be present in the user-generated tags and, thus, two different tags could be used interchangeably to annotate the same tracks (e.g., ``Rap'' and ``HipHop'').}

\changedRev{To allow interested readers to conduct further analyses of the identified clusters on a microgenre level, we release the full list of the top 20 tracks (and corresponding artists) per cluster, and we include---for each track and artist---all microgenre annotations (cf.~``Accompanying Resources'').}

\subsection{User Modeling and Music Track Recommendation} \label{sec:um_rs}
We build our context-aware music recommendation approach on top of a variational autoencoder (VAE) model~\cite{jordan1999introduction}. \changed{VAEs are a type of autoencoders~\cite{kramer1991nonlinear} that consist of an encoder, a decoder, and a loss function. In contrast to classic autoencoders, which learn encodings directly, VAEs learn the distribution of encodings using variational inference. Via sampling from the learned distribution, more representations of the same items can be generated given the same amount of training data. Thus, VAEs can learn more complex items than classic autoencoders.} 

\changed{We opted to extend the VAE architecture for collaborative filtering presented by~\citet{Liang:2018:VAC:3178876.3186150} because in a large-scale study conducted by \citet{10.1145/3298689.3347058}, the approach followed by \citet{Liang:2018:VAC:3178876.3186150} was found the only deep neural network-based approach that outperformed equally well tuned non-deep-learning approaches.
In addition, \citet{Liang:2018:VAC:3178876.3186150} evaluated their VAE architecture on the Million Song Dataset~\cite{DBLP:conf/ismir/Bertin-MahieuxEWL11}, a common benchmark in the music domain. They showed substantially superior performance compared to several baselines, in particular, the linear model weighted matrix factorization (WMF) and collaborative denoising autoencoders (CDAE).
}

As depicted in Figure~\ref{fig:vae_context_architecture}, we extend the VAE architecture by integrating context information using a gating mechanism. 
\changed{The gate output modulates the latent code in a way to incorporate context-based (country and cluster) differences of users. The abstract concepts are weighted based on how important the models deem them for a specific user group.}\footnote{\changed{We also run experiments in which we simply concatenate track listening history and context information, but this did not show improvements over the VAE based on just the listening history.}}
Specifically, we model users in form of a 122,442-dimensional listening vector (i.e., $n\_tracks$), 
which represents their track listening history, together with context information. We investigate four different ways to define a user's context: (1)~the user's country, (2)~the cluster membership of the user's country, (3)~the Euclidean distances between the user's listening vector and all identified cluster centroids, and (4)~the Euclidean distances between the user's listening vector to all country centroids.

We derive context from the self-reported country of a user. For our VAE model with country context (i.e., model 1), a one-hot encoding of the 70 included countries is used, whereas for VAE with cluster context (i.e., model 2), context is determined by the user's country membership in a cluster (see Table~\ref{tab:optics_country_clusters_perp5}), resulting in a one-hot encoding of length 9. 
For the context models~3 and 4, we first calculate the cluster centroids, i.e., each track's listening events of all users belonging to a cluster are summed and then normalized by the total amount of listening events across all tracks. 
Subsequently, for each user, the Euclidean distances between the respective user's 
normalized feature vector and all cluster centroids are determined and used as context features for the VAE with cluster distances (i.e., model~3). Country distances are calculated accordingly, where each country is considered as its own cluster (i.e., model~4). Taken together, $n\_context$ is 70 in case of model~1 and model~4, and 9 in case of model~2 and model~3. 

\renewcommand{\arraystretch}{1.2} 
\begin{table*}[h!]
\centering 
\begin{tabularx}{\linewidth}{|c|X|}
\hline
\textbf{Cluster} & \textbf{Countries} \\
\hline

0 &	ES,	IT,	IS,	SI,	PT \\
1 &	BE,	NL,	CH,	SK,	CZ,	DE,	AT,	FI,	PL \\
2 &	GB,	EE,	JP \\
3 &	AU,	NZ,	US,	CA,	PH \\
4 &	CL,	CR,	IL,	UY \\
5 &	CO,	MX,	BG,	GR \\
6 &	RO,	EG,	IR,	TR,	IN \\
7 &	BR,	ID,	VN,	MY \\
8 &	LT,	LV,	UA,	BY,	RU,	MD,	KZ,	GE \\
\hline
\textcolor{gray}{-1} & \textcolor{gray}{AQ,	FR,	NO,	ZA,	IE,	MK,	AR,	HR,	RS,	BA,	HU,	TW,	DK,	HK,	SG,	CN,	KR,	PE,	TH,	SE,	PR,	VE,	GT} \\
\hline
\end{tabularx}
\caption{\label{tab:optics_country_clusters_perp5}Country clusters as determined by OPTICS with a minimum cluster size of 3, based on the output of a t-SNE visualization (perplexity of 5) on PCA-reduced country feature vectors (100 dimensions).
Countries identified as too noisy by OPTICS are represented as \changedRev{C}luster -1. 
}
\end{table*}
\renewcommand{\arraystretch}{1} 

Our recommendation approach assumes that each user can be represented by a latent $k$-dimensional multivariate Gaussian, which is sampled, weighted by gates derived from context information, and transformed with a non-linear function to reconstruct the initial track listening history (cf.~Figure~\ref{fig:vae_context_architecture}). 
As mentioned before, our VAE model without contextual features is based on the work of \citet{Liang:2018:VAC:3178876.3186150}. 
To integrate context models, we extend the VAE by adding a gating mechanism to feed in contextual information according to the four ways detailed above. 
In a two-layer feed-forward neural network, the initial feature vector is encoded first into an intermediate representation $enc1$ and then into a latent $k$-dimensional multivariate Gaussian. The mean values $\boldsymbol{\mu}$ and variance values $\boldsymbol{\sigma}$ are the outputs of the encoding network: 

\begin{align}
\boldsymbol{enc}_1 = \tanh \left( \boldsymbol{W}_{enc_1} \cdot \boldsymbol{t} \right)
\end{align}
\begin{align}
\boldsymbol{\mu} = \tanh \left( \boldsymbol{W}_{enc_\mu} \cdot \boldsymbol{enc}_1 \right)
\end{align}
\begin{align}
\boldsymbol{\sigma} = \tanh \left( \boldsymbol{W}_{enc_\sigma} \cdot \boldsymbol{enc}_1 \right)
\end{align}

We use $tanh$ as a nonlinearity for all layers in the autoencoder. Based on our experiments (see Section~\ref{sec:experimental_setup}), we set the size of $W_{enc_1}$ to $n\_tracks$ $\times$ 1,200 and both $W_{enc_\mu}$ and $W_{end_\sigma}$ to 1,200 $\times$ 600. This results in a length of 1,200 for $enc_1$ and 600 for the latent representation $z$. The user context, given by its input vector $\boldsymbol{c}$ is transformed by a dense layer with sigmoid nonlinearity into a context gate $c_{gate}$ of the same length as latent $z$. Next, the gate is applied with component-wise multiplication to $z$:

\begin{align}
\boldsymbol{c}_{gate} = \sigma \left( \boldsymbol{W}_{context} \cdot \boldsymbol{c} \right)
\end{align}
\begin{align}
\boldsymbol {\epsilon} \sim  \mathcal{N}(\boldsymbol{0},\boldsymbol{1})
\end{align}
\begin{align}
\boldsymbol{z} = \left(\boldsymbol{\mu} + \boldsymbol{\sigma} 
\odot \boldsymbol{\epsilon}\right) \odot \boldsymbol{c_{gate}}
\end{align}

The weighted latent representation is then decoded back into the original space by a network with mirroring size but different learned parameters of the encoder:

\begin{align}
\boldsymbol{dec}_1 = \tanh \left( \boldsymbol{W}_{dec_1} \cdot \boldsymbol{z} \right)
\end{align}
\begin{align}
\boldsymbol{\hat{t}} = \tanh \left( \boldsymbol{W}_{dec_2} \cdot \boldsymbol{dec}_1 \right)
\end{align}

The detailed data flow and computation in each layer is visualized in Figure~\ref{fig:vae_context_architecture}. Based on the known track history of a target user, the models generate a variational distribution $\hat{t}$. Top-$k$ track recommendations are then retrieved by ranking the mean values of this distribution.



\section{Results and Discussion}\label{sec:results}
In the following, we present and interpret the results of our approach to identify country clusters and archetypes of music listening preferences (Section~\ref{sec:results_clustering}) and of the music track recommendation experiments (Section~\ref{sec:results_recsys}). We 
further connect the discussion to the initial research questions, which we answer in the context of the obtained results.

\subsection{Clustering of countries according to music listening preferences}\label{sec:results_clustering}
We present the identified clusters and discuss the relationship of the countries subsumed in each cluster beyond music preferences \changed{(Section~\ref{sec:cluster_descriptions})}, for instance, in terms of geographic proximity, linguistic similarities, and 
historical background. 
\changed{Furthermore, we discuss differences in user characteristics such as the users' gender, age, and their listening patterns in terms of playcounts.} 
In Section~\ref{sec:cluster_musicarchetypes}, we describe the characteristics of the clusters with respect to music preferences, i.e, we detail the track preferences that characterize the corresponding music archetypes.

\subsubsection{Identified country clusters}\label{sec:cluster_descriptions}
\begin{figure}[h]
\begin{center}
\fbox{\includegraphics[width=.45\linewidth,trim=0cm 0cm 0cm .9cm,clip]{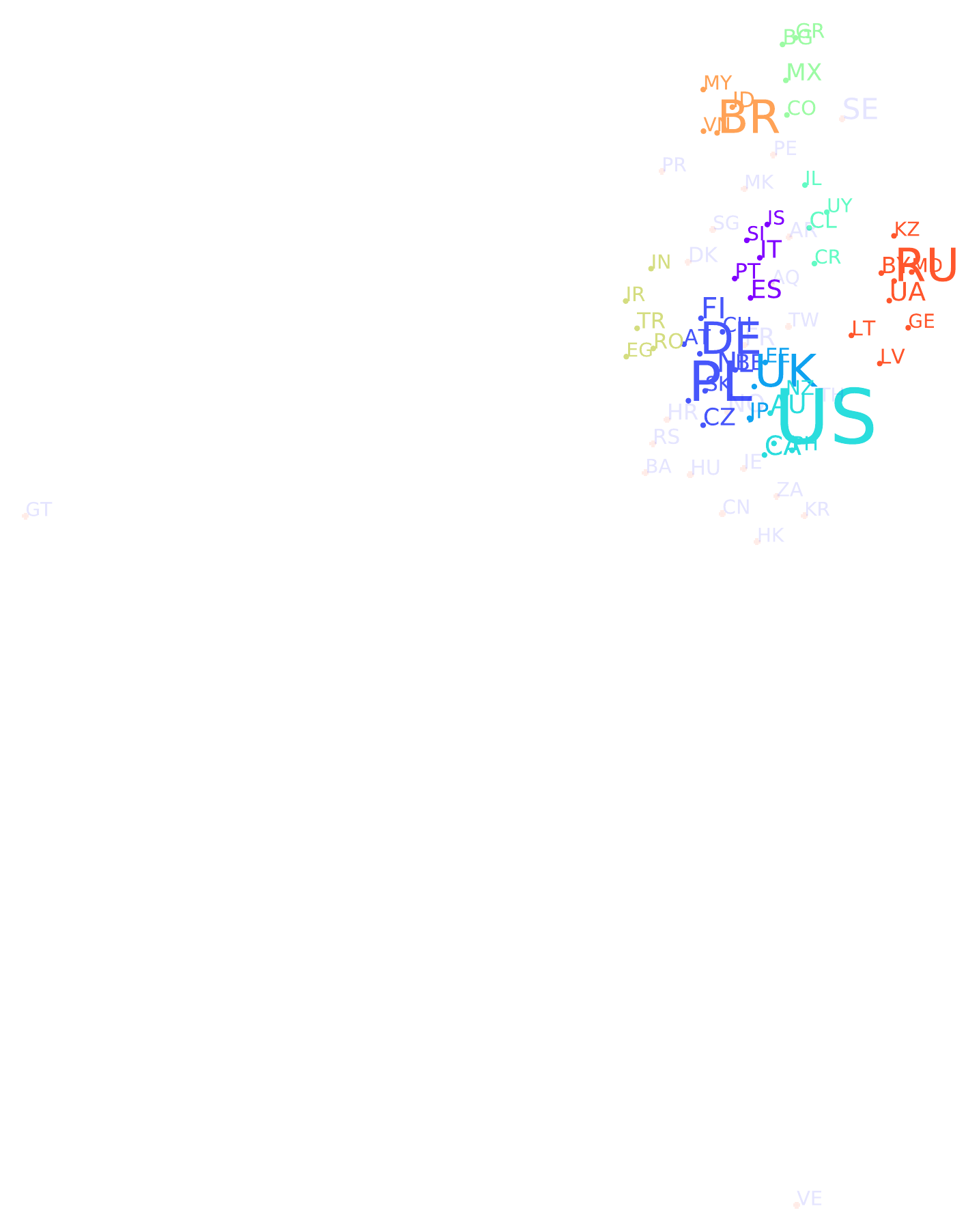}}
\fbox{\includegraphics[width=.48\linewidth,trim=0cm 0cm 0cm 0cm,clip]{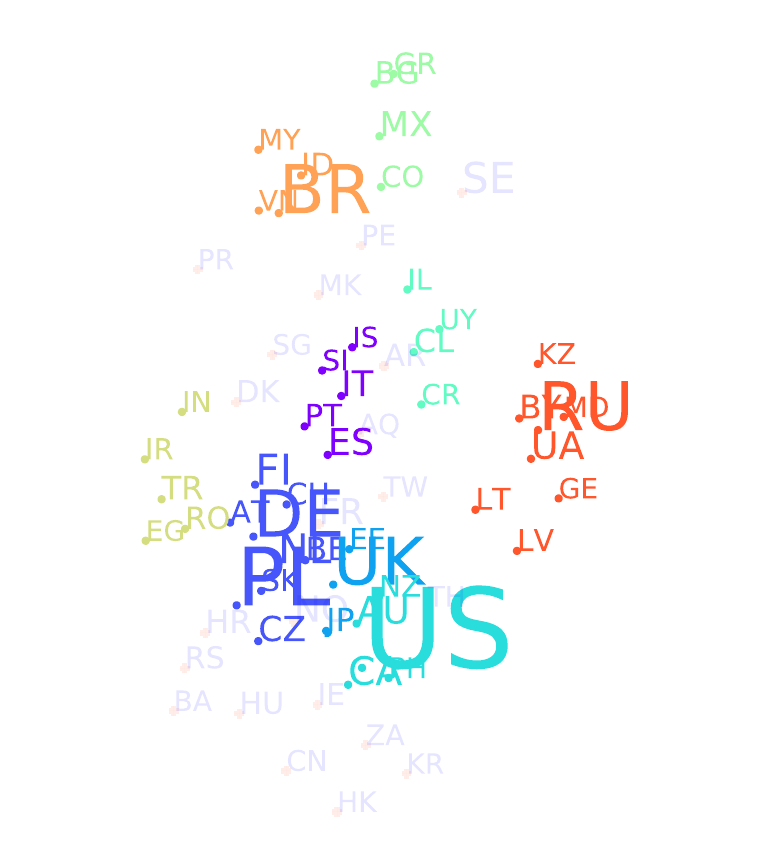}}

\end{center}
\caption{\label{fig:tsne_perp5}Results of t-SNE (perplexity of 5) and OPTICS (minimum cluster size of 3) on country feature vectors. The left part shows the full t-SNE output space, the right part a zoomed version onto the major clusters. 
}
\end{figure}

Using the approach described in Section~\ref{sec:clusters}, we can identify nine country clusters, which are presented in Table~\ref{tab:optics_country_clusters_perp5} and visualized in Figure~\ref{fig:tsne_perp5}. 
Cluster~0 contains Spain (ES), Portugal (PL), Italy (IT), Slovenia (SI), and Iceland (IS).
Cluster~1 includes as many as nine countries: Belgium (BE), The Netherlands (NL), Austria (AT), Switzerland (CH), Germany (DE), Czech Republic (CZ), Slovakia (SK), Poland (PL), and Finland (FI).
Cluster~2 refers to the United Kingdom (GB), Estonia (EE), and Japan (JP).
Cluster~3 includes Australia (AU), New Zealand (NZ), the United States (US), Canada (CA), and the Philippines (PH).
Cluster~4 refers to Chile (CL), Costa Rica (CR), Uruguay (UY), and Israel (IL).
Cluster~5 contains Colombia (CO), Mexico (MX), Bulgaria (BG), and Greece (GR).
Cluster~6 the following countries: Romania (RO), Egypt (EG), Iran (IR), Turkey (TR), and India (IN).
Cluster~7 is composed of Brazil (BR), Indonesia (ID), Vietnam (VN), and Malaysia (MY).
Cluster~8 encompasses eight countries: Lithuania (LT), Latvia (LV), Ukraine (UA), Belarus (BY), Russia (RU), Moldova (MD), Kazakhstan (KZ), and Georgia (GE).

\setlength{\fboxrule}{0.2pt}
\setlength{\fboxsep}{0pt}

Four of the countries in Cluster~0 are geographically tied together, sharing national borders (i.e., Spain (ES), Portugal (PL), Italy (IT), and Slovenia (SI)). Only Iceland (IS) is geographically dislocated. Furthermore, Spain (ES), Portugal (PL), and Italy (IT) share their roots in Romanian language. \changed{Moreover,} 
there is a Slovene minority in Italy (IT), which may lead to partly similar music preferences in Slovenia (SI) and Italy (IT).

\begin{figure}[h]
\begin{center}
\includegraphics[width=1.00\linewidth,trim=0cm 0cm 0cm 0cm,clip]{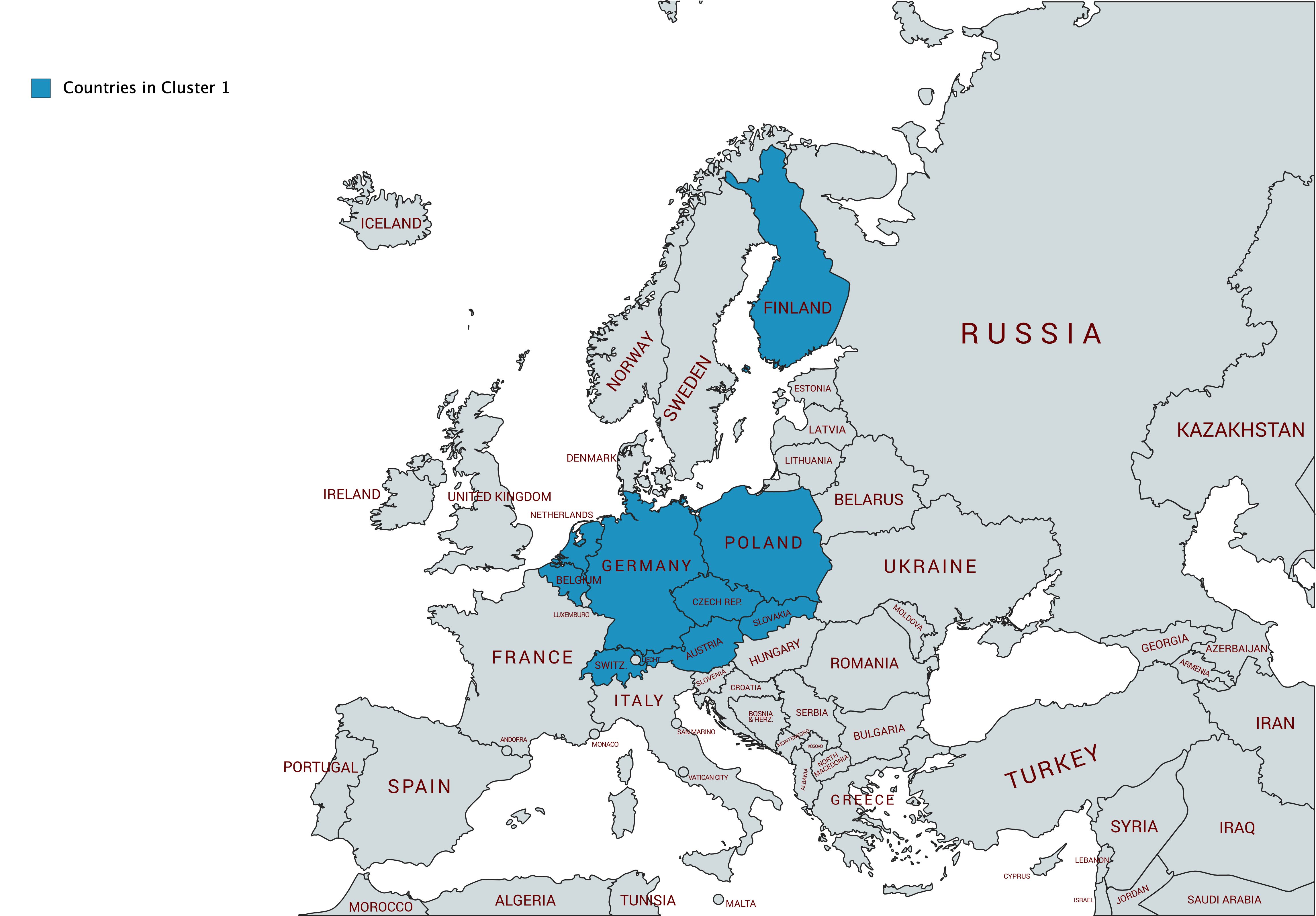}
\end{center}
\caption{\label{fig:cluster1_map}Countries in Cluster~1 \changedRev{on a map}.}
\end{figure}

Cluster~1 contains nine countries. Belgium (BE) and the Netherlands (NL) are neighboring countries and share the official language spoken (note, Belgium (BE) has two official languages). Austria (AT), Switzerland (CH), and Germany (DE) share the German language (note, Switzerland (CH) has four official languages). Czech Republic (CZ) and Slovakia (SK) are not only neighboring countries, but actually formed one joint country until 1992. 
The languages spoken in the Czech Republic (CZ), Slovakia (SK), and Poland (PL)---a neighboring country to the former two---show strong linguistic similarities.
Altogether, we can see that Belgium (BE), the Netherlands (NL), Austria (AT), Switzerland (CH), Germany (DE), Czech Republic (CZ), Slovakia (SK), and Poland (PL) are geographically connected, sharing national borders (cf.~Figure~\ref{fig:cluster1_map}). Only Finland (FI) is geographically disconnected from the other countries in this cluster.

\begin{figure}[h]
\begin{center}
\includegraphics[width=0.70\linewidth,trim=0cm 0cm 0cm 0cm,clip]{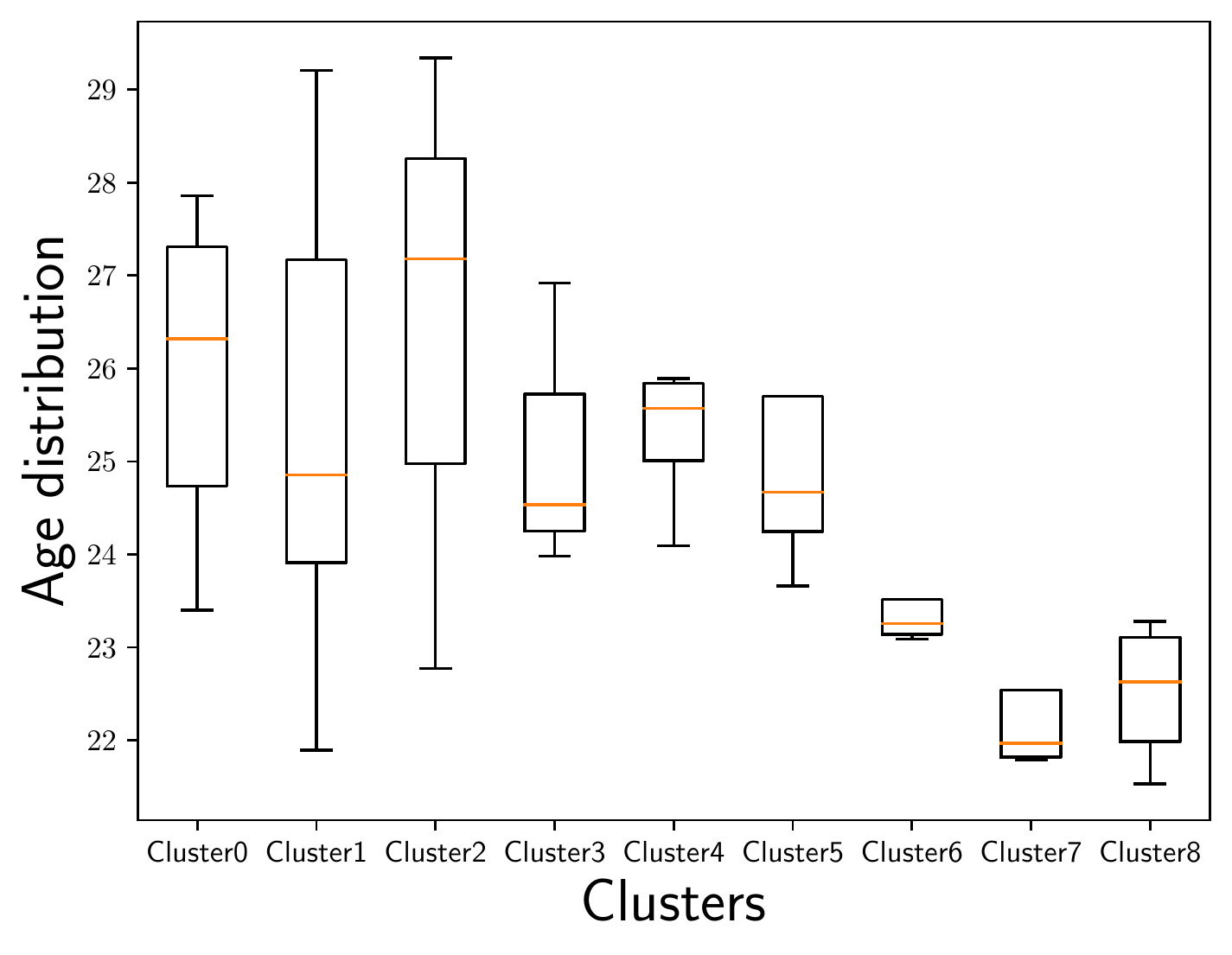}
\end{center}
\caption{\label{fig:age_clusters}\changed{Age distribution of users in the identified country clusters. While the oldest users can be found in Cluster~2, the youngest can be found in Cluster~7.}}
\end{figure}

\begin{figure}[h]
\begin{center}
\includegraphics[width=0.70\linewidth,trim=0cm 0cm 0cm 0cm,clip]{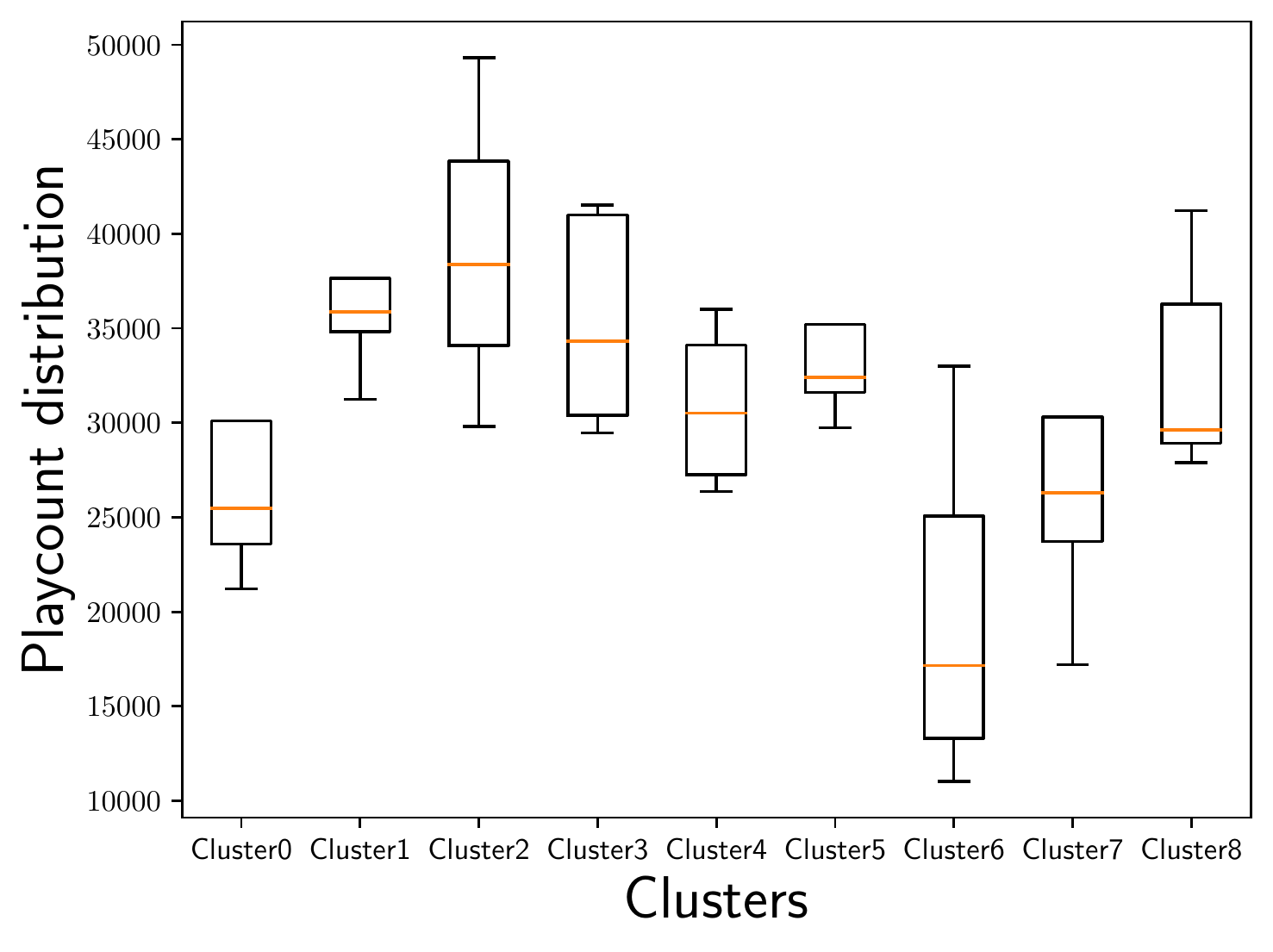}
\end{center}
\caption{\label{fig:average_playcountperuser_clusters}\changed{Distribution of users' average playcount in the identified country clusters. While the highest average playcount can be found in Cluster~2, the lowest one can be found in Cluster~6.}}
\end{figure}

Cluster~2 delivers a highly surprising result because it contains three countries that are geographically far away from each other \changed{without any linguistic similarities or close historical connections}: the United Kingdom (GB), Estonia (EE), and Japan (JP). The United Kingdom (GB) and Estonia (EE) are located at the Northwest and the Northeast of Europe---thus, at the opposite borders of Europe; Japan (JP) is even almost 8,000~km farther east of Estonia (EE). Although this cluster contains only three countries, with Japan (JP) and the United Kingdom (GB), it embraces two of the largest music markets worldwide \cite{statista_uk}. Interestingly, the United Kingdom (GB) is not part of Cluster~3 that includes most English-speaking countries. 
\changed{Considering the age distribution (Figure~\ref{fig:age_clusters}) in the identified country clusters, we find that Cluster~2 shows the highest average age with a relatively large span.\footnote{\changedRev{Please note that observations concerning age relate to our sample of Last.fm users.}} Furthermore, Cluster~2 shows by far the highest average playcount per user for the countries in this cluster (Figure~\ref{fig:average_playcountperuser_clusters}). This indicates that users in this cluster are characterized as being `power listeners'. As the combination of countries in this cluster seems surprising, age and listening intensity may be the hidden---though determining---aspects for the emergence of this cluster.}

The major connector of the countries in Cluster~3 is that they are all English-speaking countries: Australia (AU), New Zealand (NZ), United States (US), Canada (CA), and the Philippines (PH), where English is one of the two official languages in both Canada (CA) and the Philippines (PH).

Cluster~4 \changed{comprises} the countries Chile (CL), Uruguay (UY), Costa Rica (CR), and Israel (IL). Both Chile (CL) and Uruguay (UY) are located in South America and are connected by their language: Spanish. The official language in Costa Rica (CR) is Spanish as well; located in Middle America, the geographic distance to Chile (CL) and Uruguay (UY) is not far. Israel (IL), in contrast, is a country in the Middle East and is, thus, geographically disconnected from the other three countries in this cluster.

Cluster~5 contains two Latin-American countries as well as two countries in Southeastern Europe.
The Latin-American countries\changedRev{, i.e., }Mexico (MX) and Colombia (CO), are both Spanish-speaking countries. With Mexico (MX) located in the Southern part of North America and Colombia (CO) being part of South America, these are no neighboring countries, though. \changedRev{The two countries in Southeastern Europe, i.e., Bulgaria (BG) and Greece (GR), share a border. Thus, the cluster contains two country groups, which are geographically spread.} 

\begin{figure}[h]
\begin{center}
\includegraphics[width=0.70\linewidth,trim=0cm 0cm 0cm 0cm,clip]{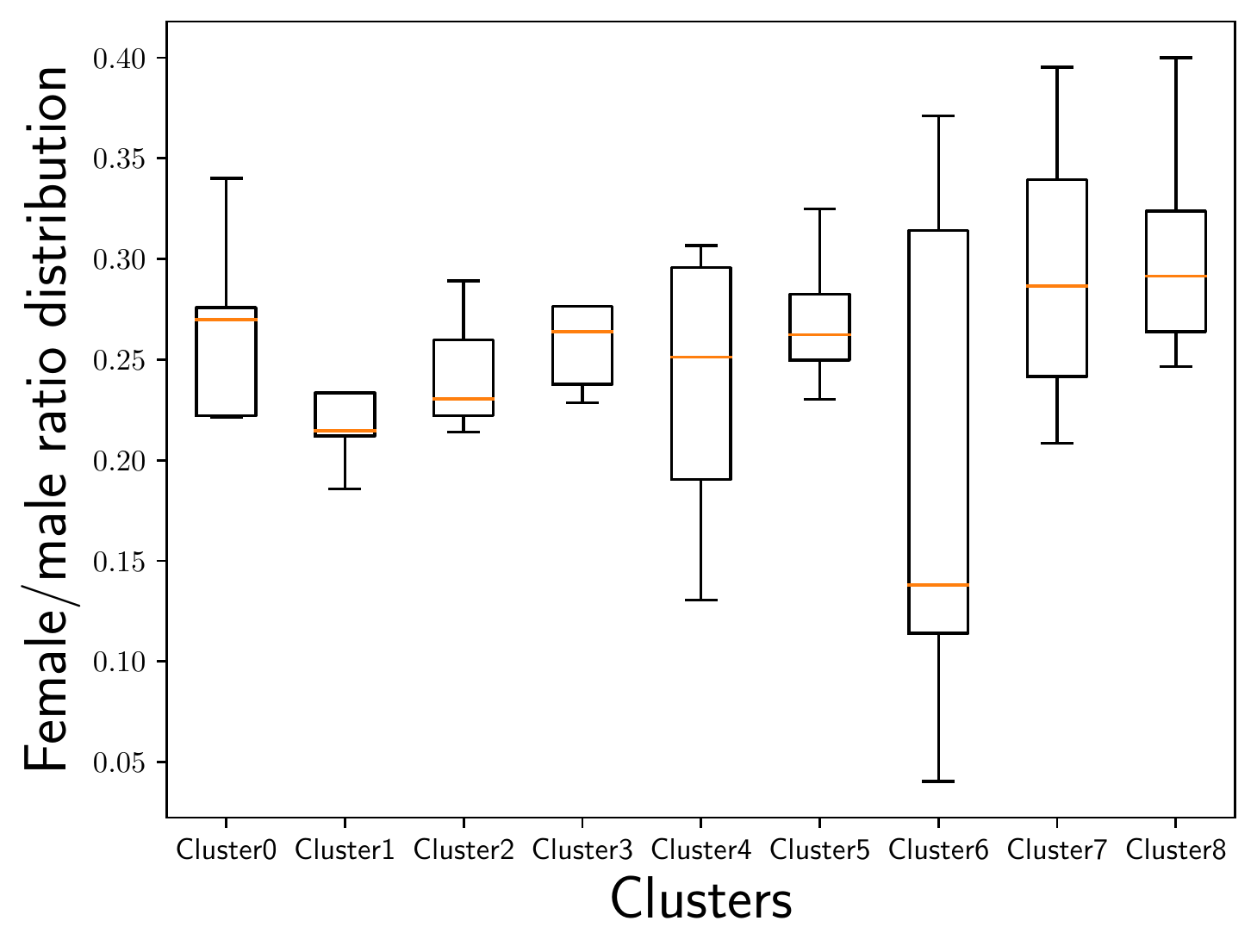}
\end{center}
\caption{\label{fig:gender_clusters}\changed{Female/male ratio distribution of users in the identified country clusters. We find that the female/male ratio is most unevenly distributed in Cluster~6 and most evenly distributed in Cluster~7.}}
\end{figure}

The countries in Cluster~6 are geographically connected, \changed{centered around countries being part of the Middle East---Turkey (TR), Iran (IR), and Egypt (EG)---and flanked by Romania (RO), that has historical relations to the others due to the Osman Empire, and India (IN), that is adjacent to the Middle East and, thus, shows a geographical proximity to the other countries in this cluster}. 
Furthermore, \changed{all the} countries in Cluster~6 are very diverse when it comes to 
\changed{the} various (minority) languages spoken, 
which may also be reflected in music preferences. \changed{Considering the female/male ratio of users (Figure~\ref{fig:gender_clusters}) in the identified country clusters, we find that Cluster~6 shows the most unevenly distributed ratio across the countries in this cluster. Despite the wide span of female/male ratios in this cluster's countries, Cluster~6 is the cluster with the overall lowest female/male ratio compared to the other clusters. With respect to age (Figure~\ref{fig:age_clusters}), this cluster comprises rather young users \changedRev{in our sample of the Last.fm community} (with the average age of users in the Clusters~7 and~8 being even younger, though). Overall, with respect to age and gender, Cluster~6 seems to have a differentiating profile compared to the other clusters. Furthermore, Cluster~6 shows by far the lowest average playcount per user (Figure~\ref{fig:average_playcountperuser_clusters}). This low number could be the result of a listening pattern that is shaped by cultural aspects, but could, for instance, also be the consequence of limited access to the resources (e.g., broadband Internet connection, streaming platforms operating in the respective countries, licenses for music content). Considering those and similar aspects is a fruitful path for future research.}

Cluster~7 \changed{covers} three neighboring countries (with maritime borders) in the Southeast of Asia---Indonesia (ID), Vietnam (VN), and Malaysia (MY)---and Brazil (BR) in South America. The three countries in the Southeast of Asia have many similarities, including common frames of reference in history, culture, and religion; also their national languages are closely related. \changed{From a geographic perspective, Brazil (BR) appears being disconnected from the other countries in this cluster.} The connection of Brazil (BR) with Indonesia (ID) and Malaysia (MY) is that all three countries have formerly been Portuguese colonies 
\cite{portuguese_colonies}. \changed{Whether this historical connection is indeed also conclusive for similar music preferences is subject to further research. Referring back to Figure~\ref{fig:age_clusters}, where we plot the age distribution for the identified country clusters, and Figure~\ref{fig:gender_clusters}, where we plot the female/male ratio, we see that Cluster~7 shows the lowest average age and is close to the highest female/male ratio. Furthermore, the female/male ratio is very evenly distributed in Cluster~7. 
We, thus, suspect that age and gender are the hidden factors construing 
this cluster or, at least, accentuating it.}

\begin{figure}[h]
\begin{center}
\includegraphics[width=1.00\linewidth,trim=0cm 0cm 0cm 0cm,clip]{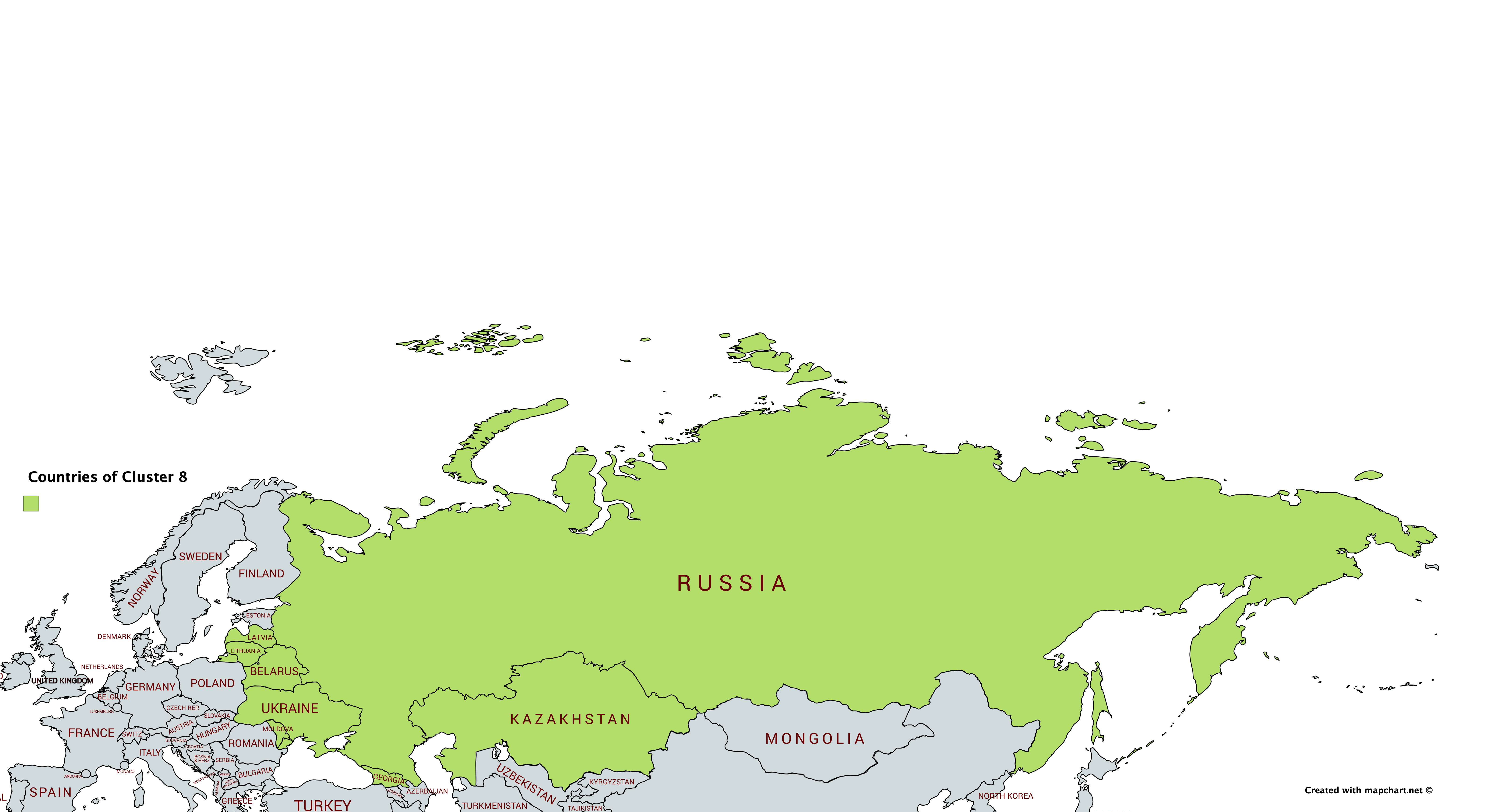}
\end{center}
\caption{\label{fig:cluster8_map}Countries in Cluster~8 \changedRev{on a map}.}
\end{figure}

As can be seen from Figure~\ref{fig:cluster8_map}, Cluster~8 comprises nine countries that are in geographical proximity: the Baltic countries Lithuania (LT) and Latvia (LV), the Russian Federation (RU), Ukraine (UA), Belarus (BY), Moldova (MD), Kazakhstan (KZ), and Georgia (GA). Besides being characterized by the geographic proximity, these countries share a history of having been part of the Russian empire. Russian is a major (or influential) language in all of the countries in this cluster
~\cite{cia_languages}.

\changed{Overall, we note that the country clusters show different characteristics with respect to age (Figure~\ref{fig:age_clusters}), gender (Figure~\ref{fig:gender_clusters}), and average playcount per user (Figure~\ref{fig:average_playcountperuser_clusters}).
With respect to age, we find especially large differences between the Clusters~2 and~7: While the highest average age can be found in Cluster~2, the lowest average age can be found in Cluster~7. The female/male ratio is high in Cluster~7 and also evenly distributed. In contrast, the female/male ratio is most unevenly distributed in Cluster~6 with a high span of ratios across the countries in this cluster; and overall, the ratio is---in comparison to the other clusters---very low.
With respect to the average playcount per user, it is also the Clusters~2 and~6 that show the largest differences: Among the users in Cluster~2 there seems to be a high ratio of `power listeners', whereas the average playcount of users in Cluster~6 is low in comparison. Overall, it can, thus, not be rejected that those and similar aspects may be hidden factors that accentuate the differentiation between the clusters or may even be indicative for the emergence of those clusters.}


\subsubsection{Characteristics of the identified clusters and music preference archetypes}\label{sec:cluster_musicarchetypes} 
To address the question what characterizes the various clusters in terms of music preferences, we use the approach described in Section~\ref{sec:clusters} to identify the most important tracks and genres for each cluster.
Table~\ref{tab:clusters_tracks} provides a list of the 10~tracks with the highest playcounts per cluster (after the IDF-based filtering explained in Section~\ref{sec:clusters}) and their genre annotations\changedRev{\footnote{\changedRev{We released the full list of the top 20 tracks (and corresponding artists) per cluster and all microgenre annotations (for each track and artist) (cf. ``Accompanying Resources'').}}}\changed{; for genre annotations, we rely on the user-generated annotations retrieved from the Last.fm~API and aligned with the Spotify microgenres, 
as described in Section~\ref{sec:clusters}}.
These most important tracks define the music preference archetypes corresponding to each cluster.

\begin{table*}[!ht]
\centering
\resizebox*{!}{\dimexpr\textheight-2\baselineskip\relax}{%
\begin{tabular}{@{}rllrl@{}}
\toprule
Cluster no. & Track title                           & Artist                        & Playcount within cluster & Track genres                        
       \\ \midrule
0          & Mr. Brightside                        & The Killers                   & 4248                     & rock,  indie rock, alternative rock          \\
           & Uprising                              & Muse                          & 3955                     & alternative rock,  rock, progressive rock    \\
           & I Bet You Look Good on the Dancefloor & Arctic Monkeys                & 3835                     & indie rock,  rock, alternative rock          \\
           & Fluorescent Adolescent                & Arctic Monkeys                & 3772                     & indie rock,  rock, alternative rock          \\
           & VCR                                   & The xx                        & 3597                     & electronic,  indie rock, indie pop           \\
           & Reptilia                              & The Strokes                   & 3394                     & indie rock,  rock, alternative rock          \\
           & Mardy Bum                             & Arctic Monkeys                & 3345                     & indie rock,  rock, alternative rock          \\
           & Hopp\'ipolla                            & Sigur Rós                     & 3336                     & post-rock,  ambient, post-rock               \\
           & There Is a Light That Never Goes Out  & The Smiths                    & 3289                     & new wave,  rock, brit-pop                    \\
           & Teardrop                              & Massive Attack                & 3260                     & triphop,  electronic, downtempo            \\
\midrule
1          & Set Fire to the Rain                  & Adele                         & 20460                    & soul,  pop, singer/songwriter               \\
           & Little Lion Man                       & Mumford \& Sons               & 20160                    & folk,  indie folk, banjo                    \\
           & Otherside                             & Red Hot Chili Peppers         & 19469                    & rock,  alternative rock, funk               \\
           & Radioactive                           & Imagine Dragons               & 19338                    & rock,  indie rock, alternative rock          \\
           & VCR                                   & The xx                        & 19220                    & electronic,  indie rock, indie pop           \\
           & Heart Skipped a Beat                  & The xx                        & 19004                    & electronic,  indie rock, rock               \\
           & Teardrop                              & Massive Attack                & 18810                    & triphop,  electronic, downtempo            \\
           & Sail                                  & AWOLNATION                    & 18728                    & electronic,  rock, indie rock               \\
           & The Pretender                         & Foo Fighters                  & 18636                    & rock,  alternative rock, grunge             \\
           & Cosmic Love                           & Florence + the Machine        & 18486                    & indie pop,  rock, pop                       \\
\midrule
2          & There Is a Light That Never Goes Out  & The Smiths                    & 7479                     & new wave,  rock, brit-pop                    \\
           & Mr. Brightside                        & The Killers                   & 7128                     & rock,  indie rock, alternative rock          \\
           & Little Lion Man                       & Mumford \& Sons               & 6979                     & folk,  indie folk, banjo                    \\
           & R U Mine?                             & Arctic Monkeys                & 6408                     & indie rock,  rock, alternative rock          \\
           & I Bet You Look Good on the Dancefloor & Arctic Monkeys                & 6302                     & indie rock,  rock, alternative rock          \\
           & I Miss You                            & blink-182                     & 6295                     & rock,  punk, pop-punk                       \\
           & Teardrop                              & Massive Attack                & 6187                     & triphop,  electronic, downtempo            \\
           & The Cave                              & Mumford \& Sons               & 6150                     & folk,  indie folk, banjo                    \\
           & VCR                                   & The xx                        & 6147                     & electronic,  indie rock, indie pop           \\
           & Harder Better Faster Stronger         & Daft Punk                     & 6083                     & electronic,  house, electronica            \\
\midrule
3          & It Ain't Cool To Be Crazy About You   & George Strait                 & 19048                    & country,  traditional country,              \\
           & Electric Feel                         & MGMT                          & 18108                    & electronic,  electronica, indie pop         \\
           & Little Lion Man                       & Mumford \& Sons               & 17089                    & folk,  indie folk, banjo                    \\
           & Time to Pretend                       & MGMT                          & 16802                    & electronic,  indietronica, electronica     \\
           & Flume                                 & Bon Iver                      & 16032                    & folk,  singer/songwriter, indie folk         \\
           & In the Aeroplane Over the Sea         & Neutral Milk Hotel            & 15753                    & indie rock,  folk, lofi                     \\
           & Midnight City                         & M83                           & 15635                    & electronic,  electro-pop, electro           \\
           & 1901                                  & Phoenix                       & 15591                    & indie pop,  electronic, indie rock           \\
           & Such Great Heights                    & The Postal Service            & 15481                    & electronic,  indie pop, electronica         \\
           & The Cave                              & Mumford \& Sons               & 15412                    & folk,  indie folk, banjo                    \\
\midrule
4          & Mephisto                              & Dead Can Dance                & 2468                     & ambient,  medieval, folk                   \\
           & 3 Libras                              & A Perfect Circle              & 1284                     & alternative rock,  progressive rock, rock    \\
           & Ariane                                & Nova                          & 1238                     & --                                         \\
           & World's End                           & Hatsune Miku \& Megurine Luka & 1228                     & --                                         \\
           & Mr. Brightside                        & The Killers                   & 1109                     & rock,  indie rock, alternative rock          \\
           & Las Fuerzas                           & Dënver                        & 1080                     & --                                         \\
           & Jeremy                                & Pearl Jam                     & 1069                     & grunge,  rock, alternative rock             \\
           & Reckoner                              & Radiohead                     & 1064                     & alternative rock,  rock, experimental       \\
           & Them Bones                            & Alice in Chains               & 1057                     & grunge,  rock, alternative rock             \\
           & Nude                                  & Radiohead                     & 1050                     & alternative rock,  rock, electronic         \\
\midrule
5          & Häaden Two                            & Robert Fripp                  & 11616                    & --                                         \\
           & The Smile                             & Phase                         & 7898                     & alternative rock,  progressive rock, art rock \\
           & Ibidem                                & Phase                         & 7858                     & alternative rock,  art rock, rock            \\
           & Perdition                             & Phase                         & 7752                     & rock,  psychedelic rock, progressive rock    \\
           & Transcendence                         & Phase                         & 7690                     & psychedelic rock,  rock, alternative rock    \\
           & Hypoxia                               & Phase                         & 7614                     & psychedelic rock,  rock, alternative rock    \\
           & Static                                & Phase                         & 6988                     & rock,  progressive rock, space rock          \\
           & A Void                                & Phase                         & 6913                     & rock,  alternative rock, indie rock          \\
           & Static (Live)                         & Phase                         & 6877                     & progressive rock,  psychedelic rock, rock    \\
           & Evening On My Dark Hillside           & Phase                         & 6793                     & psychedelic rock,  rock, alternative rock    \\
\midrule
6          & If I Could                            & Sophie Zelmani                & 13420                    & singer/songwriter,  pop, folk               \\
           & I Can't Change {[}New Song{]}         & Sophie Zelmani                & 13409                    & --                                         \\
           & Without God                           & Katatonia                     & 8024                     & doom metal,  metal, death metal              \\
           & Day                                   & Katatonia                     & 7947                     & doom metal,  metal, progressive metal        \\
           & Lady Of The Summer Night              & Omega                         & 6787                     & rock                                       \\
           & Sorrow                                & Pink Floyd                    & 6485                     & progressive rock,  rock, classic rock        \\
           & Equinoxe Part 5                       & Jean Michel Jarre             & 6457                     & ambient,  electronic rock,                  \\
           & Gammapolis                            & Omega                         & 5958                     & classic rock,  progressive rock, space rock   \\
           & To Know You                           & Sophie Zelmani                & 4783                     & singer/songwriter,  folk, pop               \\
           & To Know You (Alt. Version)            & Sophie Zelmani                & 4641                     & --                                         \\
\midrule
7          & Set Fire to the Rain                  & Adele                         & 17247                    & soul,  pop, singer/songwriter               \\
           & Fluorescent Adolescent                & Arctic Monkeys                & 13007                    & indie rock,  rock, alternative rock          \\
           & Parade                                & Garbage                       & 11770                    & rock,  alternative rock, pop                \\
           & National Anthem                       & Lana Del Rey                  & 11602                    & indie pop,  pop, triphop                    \\
           & Skyscraper                            & Demi Lovato                   & 11451                    & pop,  pop-rock, disney                      \\
           & Come \& Get It                        & Selena Gomez                  & 11387                    & pop,  electro-pop, dubstep                  \\
           & Pumped Up Kicks                       & Foster the People             & 11171                    & indie pop,  pop, indie rock                  \\
           & Dark Paradise                         & Lana Del Rey                  & 11056                    & pop,  indie pop, chamber-pop                 \\
           & Heart Attack                          & Demi Lovato                   & 10606                    & pop,  electro-pop, pop-rock                  \\
           & You Only Live Once                    & The Strokes                   & 10501                    & indie rock,  rock, alternative rock          \\
\midrule
8          & Another Bottle Down                   & Asking Alexandria             & 19779                    & post-hardcore, metal-core, screamo           \\
           & Only You                              & Savage                        & 17657                    & disco, pop, new wave                        \\
           & ...Meltdown                           & Enter Shikari                 & 16320                    & post-hardcore, trance-core, electronic       \\
           & What You Want                         & Evanescence                   & 12345                    & rock, alternative metal, gothic rock         \\
           & Gandhi Mate Gandhi                    & Enter Shikari                 & 12273                    & post-hardcore, electronic, dubstep          \\
           & Dexter                                & Ricardo Villalobos            & 11889                    & minimal, minimal techno, electronic         \\
           & Paradise Circus                       & Massive Attack                & 9922                     & triphop, electronic, downtempo             \\
           & Teardrop                              & Massive Attack                & 9891                     & triphop, electronic, downtempo             \\
           & Kill Mercy Within                     & Korn                          & 9484                     & numetal, electronic, dubstep               \\
           & Seven Nation Army                     & The White Stripes             & 9380                     & rock, alternative rock, indie rock           \\
           
\bottomrule
\end{tabular}%
}

\caption{\label{tab:clusters_tracks}The 10 most popular tracks per cluster. Playcount refers to the total number of listening events by the users in each cluster.}
\end{table*}

The most popular tracks in Cluster~0 are mainly attributed to the \changed{micro}genres indie rock and alternative rock. \changedRev{Six tracks in the top~20 have indie rock as the most associated microgenre, three~alternative rock. Eight of 20 tracks have both indie rock as well as alternative rock within their five most associated microgenres. All of the 19 tracks among the top~20 that have microgenres on track level (Si Te Quisieras Venir by the Los Planetas does not have microgenres assigned on a track level), are associated with indie rock or alternative rock; most of them with both.} Only a few tracks \changed{in later positions (thus, not in the top~10)} deviate from these genres (e.g., Set Fire to the Rain by Adele \changed{ranks on position~14 and} is associated with the genres soul and pop, Hurt by Johnny Cash \changed{is on position~16 and} is mainly associated with country and folk, or Get Lucky by Daft Punk feat.~Pharrell Williams \changed{on the position~20 that is associated with electronic)}. 
With 5 
of the 20 most frequently listened tracks in this cluster, the band Arctic Monkeys is particularly dominant in that cluster. 

While indie rock and alternative rock are represented in the most frequently listened 
tracks in Cluster~0 as well as Cluster~1, the tracks in Cluster~1 differentiate insofar from those in Cluster~0 as there is a tendency that the tracks include pop or electronic elements (e.g., VCR by The xx associated with electronic and indie rock or \changed{Cosmic Love} 
by Florence~+ the Machine). \changedRev{Four tracks in the top~20 have indie pop as the most associated microgenre, 3~electronic. Ten tracks in the top~20 have indie rock as well as alternative rock as tagged microgenres. For all tracks except Hurt by Johnny Cash and Lonely Day by System of a Down, pop is one of the tagged microgenres. Electronic is associated with 9 of the 20 tracks.}


\changedRev{In Cluster~2, two 
tracks that are most associated with folk are among the most popular tracks in the cluster (e.g., Little Lion Man or The Cave by Mumford \& Sons). Among the top~20, there are 4 tracks associated mostly with folk.} Tracks that are associated with electronic and pop (e.g., Judas by Lady Gaga) and tracks associated with triphop and electronic (e.g., Teardrop by Massive Attack) are also strongly represented. 
\changed{We recall Figure~\ref{fig:age_clusters} showing that Cluster~2 has the highest average age \changedRev{in our sample of Last.fm users}. The high average age of users in Cluster~2 and the tendency to like folk music are in line with previous research that found that folk music is more established among older users compared to younger ones~\cite{schedl_ism2017,schedl2017_kidrec}.
Yet, the results in \citet{schedl_ism2017} suggest that the preference for folk music is more prevalent for female than for male users; this seems not to be fully in line with the characteristics of Cluster~7 at first sight because the female/male ratio in Cluster~2 is not particularly high (Figure~\ref{fig:gender_clusters}). Delving deeper on the track characteristics, though, we notice that previous works considered a rather coarse-grained taxonomy of genres, whereas the work at hand considers microgenres. \changedRev{Table~\ref{tab:clusters_tracks} shows that the 10 most popular tracks in Cluster~2 reflect indie rock (4 out of 10), alternative rock (3 out of 10), and (indie) folk (2 out of 10).}
In previous work~\cite{schedl_ism2017}, alternative (rock) was associated rather with male users (typically with younger users, though). So the indie and alternative element may suggest a rather male audience.}

\changedRev{While the most listened song in Cluster~3 is associated with country (It Ain't Cool To Be Crazy About You by George Strait), this cluster} 
shows a lot of 
tracks \changedRev{that are tagged with folk} among the most popular ones for that cluster\changedRev{; 4 of the top~20 have it as their most associated microgenre}. The folk tracks are either associated with folk and the singer/songwriter genre (e.g., Flume or Holocene by Bon Iver) or are attributed to indie folk (e.g., In the Aeroplane Over the Sea by Neutral Milk Hotel). \changedRev{Eleven tracks in the top~20 are associated with electronic or electronica within the track's five most tagged microgenres.}

The most popular tracks in Cluster~4 are predominantly associated with 
progressive rock or \changed{alternative } 
rock (e.g., 3 Libras by A Perfect Circle). \changedRev{Within the top~20 of this cluster, 10 tracks are associated with some form of progressive rock and 2 with progressive metal, 14 with alternative rock, and 9 with some form of metal (i.e., progressive metal, alternative metal, doom metal, or with the gernic term metal).} An interesting deviation from the dominance of the rock genre \changedRev{is} the \changedRev{track} World's End by Hatsune Miku \& Megurine Luka, who is a vocaloid and j-pop artist. Indeed, all playcounts for that track are generated by a single user 9 from Chile (CL); thus, this track is not representative for Cluster~4. A further deviation 
is constituted by Por la Ventana by Gepe associated with the genres folk and singer/songwriter, which is listened to by more than one user.

The most popular tracks in Cluster~5 are mostly associated with the psychedelic rock genre. Interestingly, 11 of the 20 most popular tracks are by the band Phase. An exception from the strong psychedelic rock representation in this cluster is the track Slow Me Down by Anneke van Giersbergen, a track that is associated with the singer/songwriter genre, while the artist is mainly associated with alternative rock and metal, but also pop-rock.

Cluster~6 is characterized by a dichotomy of genres among the most popular tracks. On the one hand, there are tracks associated with singer/songwriter and pop (e.g., If I Could and I Can't Change by Sophie Zelmani). On the other hand, there is a strong representation of doom metal with tracks such as Without God and Day by Katatonia. Interestingly, both Sophie Zelmani as well as Katatonia are present with several songs among the most popular tracks in this cluster. \changed{Recalling the Figures~\ref{fig:age_clusters},~\ref{fig:gender_clusters}, and~\ref{fig:average_playcountperuser_clusters} that visualize the the user characteristics for the eight clusters, uneven distribution with respect to the female/male ratio and the generally low playcount per user (compared to the other clusters), and the young age of its users may be characterizing aspects for Cluster~6 that result in this heterogeneous picture with singer/songwriter and pop tracks, on the one hand, and the strong representation of doom metal, on the other. For instance, \citet{schedl_ism2017} found pop being more popular among female than male users, while it is the opposite for metal. Interestingly, the results of \citet{schedl_ism2017} (considering a global sample, also relying on data from Last.fm) suggest that the age group in which the users of Cluster~7 range, is the age group that likes pop least of all analyzed age groups, and for liking of meta this age group ranges in the middle field.}

The only cluster that includes many popular tracks associated with the pop genre is Cluster~7. Tracks include Skyscraper by Demi Lovato, Come \& Get It by Selena Gomez, and Dark Paradise by Lana Del Rey. \changedRev{Next to the generic tag pop (19 occurrences), the most mentioned microgenres among the top~20 in this cluster are poprock (16 occurrences) and indie pop (13 occurrences), followed by britpop (9), electro pop (6), dance pop (6), dream pop (4), synth pop (3), chamber pop (3), alternative pop (3), teen pop (2), art pop (2), power pop (1),  jangle pop (1), and k-pop (1).}
\changed{The high ratio of female users (Figure~\ref{fig:gender_clusters}) might be a cohesive characteristic in this cluster 
as already previous work has shown that female users are more inclined to listen to pop music than male users, in particular in the Last.fm community~\cite{schedl_ism2017,schedl2017_kidrec}.}

Cluster~8 is characterized by the post-hardcore genre. \changedRev{Seven tracks in the top~20 in this cluster are tagged with post-hardhore, five of those have it as their most tagged microgenre. Triphop (8 tracks), screamo (6 tracks), and hardcore (6 tracks) are also well represented among the top~20 in this cluster.} Popular tracks include Another Bottle Down by Asking Alexandria, ...Meltdown by Enter Shikari, and Nineteen Fifty Eight by A Day to Remember. An interesting deviation from this post-hardcore association are, for instance, Dexter by Ricardo Villalobos (minimal techno) and Cookie Thumper! by Die Antwoord (hip hop), which are also among the most popular tracks in this cluster.

Summarizing the answer to RQ1, which we addressed here (To what extent can we identify and interpret groups of countries that constitute \textit{music preference archetypes}, from behavioral traces of users' music listening records?), we find nine clusters of countries, with each of the clusters representing a \textit{music preference archetype} that reflects different nuances of music preferences 
in terms of the Spotify microgenres. While some \textit{music preference archetypes} represent countries with geographical proximity (e.g., Cluster~6 and Cluster~8) and some archetypes share linguistic similarities (e.g., Cluster~3 and Cluster~8), others include interesting outliers (e.g, Iceland (IS) in Cluster~0, Israel (IL) for Cluster~4, or Brazil (BR) in Cluster~7). 


\subsection{Music track recommendation using country context}\label{sec:results_recsys}
In the following, we first detail the setup of the conducted evaluation experiments for the music track recommendation task, including evaluation protocol, baselines, and performance metrics (Section~\ref{sec:experimental_setup}). Subsequently, we report and discuss the obtained results ans answer the related research questions (Section~\ref{sec:recsys_results_discussion}).

\subsubsection{Experimental setup}\label{sec:experimental_setup}
After preselection and filtering (cf.~Section~\ref{sec:data}), the dataset contains the listening histories of 54,337 Last.fm users. To carry out the recommendation experiments, we split the data into training, validation, and test sets. For each of validation and test set, 5,000 users are randomly sampled. The original VAE model~\cite{Liang:2018:VAC:3178876.3186150} and our extended VAE architecture that integrates the user context models described in Section~\ref{sec:um_rs} are trained on the full listening events of the uses in the training set. For users in the validation and test set, 80\% of all listening events are randomly selected to act as an input for the model, and the remaining 20\% are used for evaluation.
The NDCG@100 metric (see below) on the validation set is used to select the hyperparameters of our models. 

\textbf{Baselines:}
In addition to comparing our extended context-aware model to the original VAE recommendation architecture~\cite{Liang:2018:VAC:3178876.3186150}, we also include two baselines in the experiments, i.e.,~\changed{variants of} most popular (MP) and implicit matrix factorization (IMF).
In the \textit{most popular (MP)} models, a popularity measure is calculated for each track based on its sum of listening events across users in the training set. 
\changed{We implemented and evaluated three flavors of MP: \textit{MP global} computes the most popular tracks on a global scale (independent of country); \textit{MP country} considers only the top tracks in the country of the target user; \textit{MP cluster} considers only the top tracks within the cluster the country of the target user belongs to.}
We then rank tracks accordingly and use the ranking to produce recommendations, which are evaluated on the 20\% split of the test set (for each user). To make results between the baseline and our proposed model comparable, we exclude tracks that are part of a user's known listening history, i.e., listening events from the remaining 80\%.
As a second baseline, we adopt a collaborative filtering approach using \textit{implicit matrix factorization (IMF)} according to \citet{Koren:2009:MFT:1608565.1608614}.
We use the implementation provided by Spotlight\footnote{\url{https://maciejkula.github.io/spotlight/factorization/implicit.html}} with random negative sampling (50:50), 128 latent dimensions, and a pointwise loss function. 

\textbf{Performance metrics:}
To quantify the accuracy of the recommendations, we use the following metrics~\cite[similar to][]{Liang:2018:VAC:3178876.3186150,Schedl2018,Aiolli:2013:ETR:2507157.2507189}, which we report averaged over all users (in the test set).
\changed{Thus, for each user in the test set, we generate recommendations using the data in the training set and compare the recommended tracks with the actually listened tracks of the user present in the test set in order to calculate the performance metrics.} 
Note that we use the definitions common in recommender systems research, which are partly different from the ones traditionally used in information retrieval.

\changed{
\textit{Precision@K} for user $u$:\\
\begin{align}
P@K(u) = \frac{1}{K} \sum_{i=1}^{K} {rel(i)},
\end{align}
where $K$ is the number of recommended items and $rel(i)$ is an indicator function signaling whether the recommended track at rank $i$ is relevant to $u$ or not. This means that $rel(i)=1$ if the recommended track at rank $i$ can be found in the test set; $rel(i)=0$ if not. 
}

\changed{
\textit{Recall@K} for user $u$:\\
\begin{align}
R@K(u) = \frac{1}{\min(K,N_u)} \sum_{i=1}^{K} {rel(i)}
\end{align}
where $N_u$ is the number of items in the test set that are relevant to $u$, $K$ is the size of the recommendation list, and $rel(i)$ is the same indicator function as used for \textit{Precision@K}. When comparing \textit{Precision@K} and \textit{Recall@K}, \textit{Precision@K} can be seen as a measure of the usefulness of recommendations and \textit{Recall@K} as a measure of the completeness of recommendations.
}

\changed{
\textit{Normalized discounted cumulative gain@K}:\\
\begin{align}
NDCG@K(u) = \frac{DCG@K(u)}{IDCG@K(u)}
\end{align}
where $IDCG@K(u)$ is the ideal $DCG@K$ for user~$u$, achieved when all items relevant to $u$ are ranked at the top, and $DCG@K(u)$ is the \textit{discounted cumulative gain} at position $k$ for user $u$. It is given by:
\begin{align}
DCG@K(u) = \sum_{i=1}^{K} \frac{rel(i)}{\log_2 \left(i+1\right)}
\end{align}
where $rel(i)$ is the same indicator function as used for \textit{Precision@K} and \textit{Recall@K}. 
In contrast to those two performance metrics, \textit{NDCG@K} is a ranking-based metric, which also takes the position of the recommended 
tracks into account since higher-ranked items are given more weight.
}

We compute and report all metrics for $K=10$ and $K=100$, 
simulating users who are just interested in a few top recommendations and users who inspect a large part of the recommendation list, respectively.

\subsubsection{Results and discussion}\label{sec:recsys_results_discussion}

Table~\ref{tab:result_table} shows the performance 
achieved on the test set, averaged over all users in the test set.
As a general observation, we see that the VAE-based approaches outperform the baselines (MP and IMF) by a substantial extent.
\changed{Of the baselines, IMF performs superior to MP global while the other two variants of MP (MP country and MP cluster) yield better results than IMF.}
The poor performance of MP global is somewhat surprising since several studies \cite[e.g.,][]{10.1007/978-3-319-95171-3_51,DBLP:journals/ijmms/LaiLH19,Vall2019} have shown that recommendation approaches leveraging popularity information---e.g., always suggesting the items that are most frequently consumed---often achieve highly competitive accuracy values in offline experiments, despite the obvious fact that such recommendations will likely not be perceived very useful by the users.
A likely reason is that we perform track recommendation while the earlier mentioned works commonly adopt an artist recommendation setup. 
In an artist recommendation scenario, it is very likely that a user has consumed every highly popular artist at least once. This leads to a high performance of a popularity-based approach. 
In the track recommendation scenario adopted in the work at hand, the granularity of items (tracks vs.~artists) is higher and---in comparison to the artist recommendation scenario---it is not necessarily the case that the most popular tracks have been consumed by most users at least once. Overall, a popularity-based approach may work well for artist recommendation but less so for the more fine-grained track recommendation.

\renewcommand{\arraystretch}{1.2} 
\begin{table*}[!ht]
\centering 
\begin{tabularx}{\linewidth}{X|r|r|r|r|r|r}
\toprule
\textbf{Model} & \textbf{P@10} & \textbf{P@100} & \textbf{R@10} & \textbf{R@100} & \textbf{NDCG@10} & \textbf{NDCG@100}  \\
\midrule
\changed{MP global}			&	0.048 & 0.033 & 0.048 & 0.036 & 0.050 & 0.037 \\
\changed{MP country}				&	0.203 & 0.156 & 0.203 & 0.157 & 0.209 & 0.166 \\
\changed{MP cluster}				&	0.193 & 0.149 & 0.193 & 0.149 & 0.199 & 0.158 \\
IMF 	    &   0.080 & 0.072 & 0.080 & 0.064 & 0.081 & 0.071 \\
\midrule
VAE 		&	0.482 & 0.309 & 0.486 & 0.367 & 0.500 & 0.383 \\
\midrule
VAE country id (model 1)	&	0.513 & 0.325 & 0.517 & 0.384 & 0.532 & 0.402 \\
VAE cluster id (model 2)	&	0.515 & 0.326 & 0.520 & 0.385 & 0.537 & 0.404 \\
VAE cluster dist (model 3) 	&	0.513 & 0.325 & 0.518 & 0.384 & 0.534 & 0.403 \\
VAE country dist (model 4) &	0.516 & 0.325 & 0.521 & 0.383 & 0.535 & 0.403 \\
\midrule
\midrule
\changed{VAE sampling}				&	0.224 & 0.099 & 0.239 & 0.255 & 0.252 & 0.223 \\
\changed{VAE sampling country id (m.~1)}	&	0.230 & 0.102 & 0.245 & 0.259 & 0.259 & 0.227 \\
\changed{VAE sampling cluster id (m.~2)}	&	0.231 & 0.101 & 0.246 & 0.259 & 0.261 & 0.227 \\
\changed{VAE sampl.~cluster dist (m.~3)} 	&	0.232 & 0.102 & 0.245 & 0.258 & 0.246 & 0.258 \\
\changed{VAE sampl.~country dist (m.~4)} 	&	0.225 & 0.100 & 0.239 & 0.255 & 0.255 & 0.223 \\
\bottomrule
\end{tabularx}
\caption{\label{tab:result_table} Results with respect to Precision@K, Recall@K, and NDCG@K metrics. For all metrics, pairwise comparison using a Wilcoxon signed-rank test shows significant improvements from MP global to IMF to VAE to all VAE models with context (models~1--4); there are no significant differences between the 4 VAE models that use context, though.
\changed{The five rows at the bottom (``VAE sampling ...'') show results for another set of experiments in which we randomly sampled (three times) exactly 122,442 tracks from about 1 million tracks instead of computing performance measures on the top 122,442 tracks of the whole collection as done in the main experiment. 
}
}
\end{table*}
\renewcommand{\arraystretch}{1} 

\changed{On the other hand, we also note that the other two variants (MP country and MP country) achieve much better results than MP global, even outperforming the IMF approach. This might be explained by the more narrow but better user-tailored consideration of the country-specific mainstream (cf.~\citet{10.1371/journal.pone.0217389}), which is reflected in the computation of most popular tracks in the MP country and MP cluster models.}

Comparing the proposed context-aware extensions of the VAE recommendation architecture to the original VAE~\cite{Liang:2018:VAC:3178876.3186150}, we observe a clear improvement of all metrics when integrating the user context models. This improvement is achieved irrespective of the actual user model we adopt (models 1--4). 
Precision@10 increases by 3.4 percentage points (7.1\%) from VAE to the best performing VAE context model (model~4) that leverages the distances between users and country centroids. 
Likewise, Precision@100 increases by 1.7 percentage points (5.5\%).
Recall@10 and Recall@100 improve, respectively, by a maximum of 3.5 percentage points (7.2\%), realized by model~4, and by 1.8 percentage points (4.9\%), realized by model~2.
In terms of NDCG, the largest gains are realized by VAE context model~2 that incorporates cluster ids. NDCG@10 improves by 3.7 percentage points (7.4\%) compared to VAE; NDCG@100 increases by 2.1 percentage points (5.5\%).


We investigate statistical significance of the results as follows.
For all used metrics (i.e., P@10, P@100, R@10, R@100, NDCG@10, NDCG@100), data is non-normally distributed (Kolmogorov-Smirnov test, $p\le0.001$). Accordingly, we use the Friedman test \cite{Friedman1937} 
to test the models' performances for differences. For each metric, the models differ at a significance level of $p\le0.001$.
In pairwise comparisons using a Wilcoxon signed-rank test \cite{wilcoxon}, for each metric, the tests indicate that VAE outperforms each of the baselines (i.e, MP an IMF) at a significance level of \changedRev{$p\le0.05$}.
Furthermore, we perform a pairwise comparison, again using Wilcoxon's signed-rank test, for each metric and each pair of pure VAE and one of the models integrating context information (i.e., models~1--4). For each metric and each of the models~1--4, the models~1--4 outperform the pure VAE (without context integration) at a significance level of \changedRev{$p\le0.05$}.
Yet, the Friedman test did not indicate any significant differences of the models~1--4 for any of the metrics.

\vspace{4mm}

Returning to the original research questions, we answer RQ2 (Which are effective ways to model the users' geographic background as a contextual factor for music recommendation?) by pointing to the fact that all four user models proposed are effective to significantly improve recommendation quality in terms of precision, recall, and NDCG measures. We note, however, that performance differences between the four \changed{user context} models are largely negligible. In summary, leveraging country information for music track recommendation (either as country or cluster identifier, or as distances between the target user and each cluster's centroid) is beneficial compared to not including any country information.

As for RQ3 (How can we extend a state-of-the-art recommendation algorithm to include user context information, in particular, our geo-aware user models?), we proposed an extension of a state-of-the-art recommender based on a VAE architecture~\cite{Liang:2018:VAC:3178876.3186150}, i.e., we devised a multi-layer generative model in which contextual features can influence recommendations through a gating mechanism.

\vspace{5mm}
\changed{
To investigate the generalizability of our findings to a dataset with different characteristics, we perform an additional experiment as follows.
We estimate performance on a more diverse dataset in terms of track popularity than the one that considers only the top 122,442 tracks.
More precisely, we create a second dataset by first considering all tracks that have been listened to as least 100 (instead of 1,000) times, yielding 1,012,961 unique tracks. We then randomly sample, three times, exactly the same amount of tracks (122,442) as used in our main experiment, and we evaluate the VAE approaches on each randomly sampled subset, averaging performance measures across the three runs.\footnote{\changed{Please note that computational limitations prevented us from running experiments on all 1,012,961 tracks, even more so on the entire LFM-1b dataset.}}
Results can be found in the five last rows of Table~\ref{tab:result_table} (models named ``VAE sampling ...''). While we observe an obvious decrease in performance when considering items further down the popularity scale, results are still in line with the findings obtained on the main dataset. In particular, our extended VAE models (models 1--4) still outperform the original VAE architecture, with respect to all performance metrics.
}

\section{Conclusions, Limitations, and Future Work}\label{sec:conclusions}
In \textbf{summary}, 
we approached the task of identifying country clusters and corresponding archetypes of music consumption preferences based on behavioral data of music listening that originates from Last.fm users. Together with the users' self-disclosed country information, we used the listening data (369 million listening events created by 54 thousand Last.fm users) as an input to unsupervised learning techniques (t-SNE and OPTICS), allowing us to \textit{identify nine archetypal country clusters}. We discussed these clusters in detail with respect to their corresponding users' music preferences on the track level and the linguistic, historical, and cultural backgrounds of the countries in each cluster.
\changed{Additionally, we considered the distribution of age, gender, and average playcount per user as aspects in our analysis.}

Furthermore, we proposed a context-aware music recommendation approach operating on the music track level, which integrates different user models that are based on the user's country or country cluster. To this end, we \textit{extended a variational autoencoder (VAE) architecture by a gating mechanism to add contextual user features}. 
We considered \textit{four user models}, either encoding the target user's country information (model 1) or cluster information (model 2) directly, or as a feature vector containing the distances between the target user and all cluster centroids (model 3) or all individual country centroids (model 4).
In evaluation experiments, using precision, recall, and NDCG as performance metrics, we showed that all VAE architectures outperformed a popularity-based recommender and implicit matrix factorization, which served as our baselines. Results further revealed \textit{superior performance of all VAE variants that include context information} vis-\`a-vis VAE without context information, regardless of how country information is encoded in the user model. 

Yet, this work has potential \textbf{limitations} with respect to the underlying dataset, which we discuss in the following.
\changed{There are social patterns that define how and why people access music~\cite{LOPEZSINTAS201456}. A dataset containing logs of the interactions with an online platform can, thus, only capture those listening events of people using any form of online music platform. According to~\citet{LOPEZSINTAS201456}, music access patterns are structured by an individual's social position (indicated by education) and life stage (indicated by age). A bias with respect to the users' social background can therefore be expected for our dataset.}
For instance, the dataset has a strong \textit{community bias} towards users in the United States (US), while other countries are less represented. 
Furthermore, \textit{user information is self-reported} by the users, which may be prone to errors and may not necessarily reflect the truth. For instance, some users report as their country Antarctica (AQ) or a birth year of 1900, which both do not seem overly plausible---especially in combination \cite[also see Figure~1 in][]{Schedl2017}.
Moreover, some users show \textit{very high playcounts for single tracks}, which are not popular among other users. This also affects six of the tracks presented in our discussion of the music preference archetypes. For instance, World's End by Hatsune Miku \& Megurine Luka has a playcount of 1,228 generated by a single unique user. Similarly, One Thing' by Runrig and Resemnare by Valeriu Sterian both have exactly one unique listener, who generated a playcount of 4,000 and 3,591, respectively. The track Ariane by Nova has 3 unique users; 
I Can't Change [New Song] and To Know You (Alt. Version)---both by Sophie Zelmani---have 5 unique users each, whereof almost all playcounts were generated by only one single user. For both songs, this is the same user. 
\changed{Notably, also \changedRev{the preferences of the} Last.fm users \changedRev{in our dataset} towards certain genres differ from the genre preferences of the population at large. For instance, we found that rap and R\&B as well as classical music is substantially underrepresented in Last.fm listening data~\cite{DBLP:conf/mm/SchedlT14}\changedRev{, which we use in the present study}. }
To some extent, these limitations related to the dataset could be alleviated in the future by performing further data cleansing and preprocessing steps, e.g., threshold-based filtering of exorbitant playcounts by a minority of listeners. 

\changed{Another limitation of the work concerns a characteristic of t-SNE, which is that the \textit{cost function t-SNE uses is non-convex}. This, in turn, may result in a different embedding of data points in the low-dimensional output space when the t-SNE algorithm is run on different software or hardware configurations.\footnote{\changed{Note that our results are stable for a given machine, software configuration, and parameter setting since we fixed the seed of the random number generator. Running the code on other configurations, however, may result in a slightly different visualization and clustering.}}
Please note that this does not only concern the present work, but potentially the entire (large) body of research that employs t-SNE for visualization. It is, however, an aspect that is barely discussed.
We address this issue in the current work by providing exact details on our implementation and used software, and by releasing to the public the source code, parameter configurations, and dataset used in our experiments (cf. ``Accompanying Resources'' below).
}

In this work, we used simple mechanisms to integrate country information as context factors into a VAE architecture. While they worked out well, i.e., outperformed a non-context-aware VAE, we expect even better performance with other user models, whose creation will be part of \textbf{future research}.
For instance, we contemplate using \textit{probabilistic models} to describe the likelihood of each user to belong to each cluster (or country), e.g., via Gaussian mixture models.
Given the actual country of a user, we could then analyze in more detail users whose stated country 
is not the country with highest probability.
Such a framework could also be used to \textit{diversify recommendations} according to a user-selected country, fulfilling user intents such as ``I want music of my preferred genre, but listened to by Brazilians''.

Furthermore, it would be worthwhile to \textit{compare the clustering and recommendation results} we achieved here on the track level to results achieved when modeling music preferences on the artist level, keeping all other methodological details the same.
In particular, since previous studies have predominantly shown that popularity-based music recommendation systems perform well when recommending artists, such a comparison could be enlightening.


Finally, we aim at delving into the possible \textit{cultural, historical, or socio-economic reasons} that may underlie the differences in music preferences between the \changed{identified} 
archetypes. To this end, we will consider theories and insights from cultural sciences, history, sociology, and economics, and connect our music \changed{preference} archetypes 
to these theories.
\changed{Another promising path for further analysis of the country clusters is to consider dimensions rooted in the music market or the music content itself, including considerations such as as local demand, production of music styles, reception of music styles, diffusion, etc., as well as dimensions related to the users' listening habits.}

\section*{Accompanying Resources}\label{sec:resources}
To foster reproducibility, we release the code and data used in this work to the public. The code can be found on \url{https://gitlab.cp.jku.at/markus/fiai2020_country_clusters}; the dataset~\cite{schedl_markus_2020_3832071} is available from \url{https://zenodo.org/record/3907362#.XvRq1CgzZPY}. 
In our experiments, we used the following setup: Windows 10 Home, Build 18362; Python 3.7.0; T-SNE and OPTICS implementations of scikit-learn 0.21.3. The remaining, system-independent, configuration details can be found in the code.




\section*{Funding}
This research is supported by the Austrian Science Fund (FWF): V579 and the Know-Center GmbH (FFG COMET funding).

\section*{Acknowledgments}
The authors would like to thank Peter M\"ullner from the Know-Center GmbH for providing the IDF calculations of the music tracks.

\bibliographystyle{ACM-Reference-Format}
\bibliography{MusicAI}


\begin{thebibliography}{83}


\ifx \showCODEN    \undefined \def \showCODEN     #1{\unskip}     \fi
\ifx \showDOI      \undefined \def \showDOI       #1{#1}\fi
\ifx \showISBNx    \undefined \def \showISBNx     #1{\unskip}     \fi
\ifx \showISBNxiii \undefined \def \showISBNxiii  #1{\unskip}     \fi
\ifx \showISSN     \undefined \def \showISSN      #1{\unskip}     \fi
\ifx \showLCCN     \undefined \def \showLCCN      #1{\unskip}     \fi
\ifx \shownote     \undefined \def \shownote      #1{#1}          \fi
\ifx \showarticletitle \undefined \def \showarticletitle #1{#1}   \fi
\ifx \showURL      \undefined \def \showURL       {\relax}        \fi
\providecommand\bibfield[2]{#2}
\providecommand\bibinfo[2]{#2}
\providecommand\natexlab[1]{#1}
\providecommand\showeprint[2][]{arXiv:#2}

\bibitem[\protect\citeauthoryear{Adiyansjah, Gunawan, and Suhartono}{Adiyansjah
  et~al\mbox{.}}{2019}]%
        {ADIYANSJAH201999}
\bibfield{author}{\bibinfo{person}{Adiyansjah}, \bibinfo{person}{Alexander A~S
  Gunawan}, {and} \bibinfo{person}{Derwin Suhartono}.}
  \bibinfo{year}{2019}\natexlab{}.
\newblock \showarticletitle{Music Recommender System Based on Genre using
  Convolutional Recurrent Neural Networks}.
\newblock \bibinfo{journal}{\emph{Procedia Computer Science}}
  \bibinfo{volume}{157} (\bibinfo{year}{2019}), \bibinfo{pages}{99--109}.
\newblock
\showISSN{1877-0509}
\urldef\tempurl%
\url{https://doi.org/10.1016/j.procs.2019.08.146}
\showDOI{\tempurl}
\newblock
\shownote{The 4th International Conference on Computer Science and
  Computational Intelligence (ICCSCI 2019): Enabling Collaboration to Escalate
  Impact of Research Results for Society.}


\bibitem[\protect\citeauthoryear{Adomavicius and Tuzhilin}{Adomavicius and
  Tuzhilin}{2015}]%
        {adomavicius_tuzhilin:rsh:2015}
\bibfield{author}{\bibinfo{person}{Gediminas Adomavicius} {and}
  \bibinfo{person}{Alexander Tuzhilin}.} \bibinfo{year}{2015}\natexlab{}.
\newblock \showarticletitle{{Context-Aware Recommender Systems}}.
\newblock In \bibinfo{booktitle}{\emph{{Recommender Systems Handbook}}
  (\bibinfo{edition}{2nd} ed.)}, \bibfield{editor}{\bibinfo{person}{Francesco
  Ricci}, \bibinfo{person}{Lior Rokach}, \bibinfo{person}{Bracha Shapira},
  {and} \bibinfo{person}{Paul~B. Kantor}} (Eds.).
  \bibinfo{publisher}{Springer}, \bibinfo{pages}{191--226}.
\newblock


\bibitem[\protect\citeauthoryear{Aiolli}{Aiolli}{2013}]%
        {Aiolli:2013:ETR:2507157.2507189}
\bibfield{author}{\bibinfo{person}{Fabio Aiolli}.}
  \bibinfo{year}{2013}\natexlab{}.
\newblock \showarticletitle{{Efficient Top-n Recommendation for Very Large
  Scale Binary Rated Datasets}}. In \bibinfo{booktitle}{\emph{Proceedings of
  the 7th ACM Conference on Recommender Systems}} (Hong Kong, China)
  \emph{(\bibinfo{series}{RecSys '13})}. \bibinfo{publisher}{ACM},
  \bibinfo{address}{New York, NY, USA}, \bibinfo{pages}{273--280}.
\newblock
\showISBNx{978-1-4503-2409-0}
\urldef\tempurl%
\url{https://doi.org/10.1145/2507157.2507189}
\showDOI{\tempurl}


\bibitem[\protect\citeauthoryear{Ankerst, Breunig, Kriegel, and Sander}{Ankerst
  et~al\mbox{.}}{1999}]%
        {Ankerst:1999:OOP:304182.304187}
\bibfield{author}{\bibinfo{person}{Mihael Ankerst}, \bibinfo{person}{Markus~M.
  Breunig}, \bibinfo{person}{Hans-Peter Kriegel}, {and}
  \bibinfo{person}{J\"{o}rg Sander}.} \bibinfo{year}{1999}\natexlab{}.
\newblock \showarticletitle{OPTICS: Ordering Points to Identify the Clustering
  Structure}. In \bibinfo{booktitle}{\emph{{Proceedings of the 1999 ACM SIGMOD
  International Conference on Management of Data}}} (Philadelphia,
  Pennsylvania, USA) \emph{(\bibinfo{series}{SIGMOD '99})}.
  \bibinfo{publisher}{ACM}, \bibinfo{address}{New York, NY, USA},
  \bibinfo{pages}{49--60}.
\newblock
\showISBNx{1-58113-084-8}
\urldef\tempurl%
\url{https://doi.org/10.1145/304182.304187}
\showDOI{\tempurl}


\bibitem[\protect\citeauthoryear{Bada}{Bada}{2018}]%
        {portuguese_colonies}
\bibfield{author}{\bibinfo{person}{Ferdinand Bada}.}
  \bibinfo{year}{2018}\natexlab{}.
\newblock \showarticletitle{Former Portuguese Colonies}.
\newblock \bibinfo{journal}{\emph{WorldAtlas}} (\bibinfo{year}{2018}).
\newblock
\newblock
\shownote{{https://worldatlas.com/articles/former-portuguese-colonies.html}.}


\bibitem[\protect\citeauthoryear{Baek}{Baek}{2015}]%
        {baek2015}
\bibfield{author}{\bibinfo{person}{Young~Min Baek}.}
  \bibinfo{year}{2015}\natexlab{}.
\newblock \showarticletitle{Relationship Between Cultural Distance and
  Cross-Cultural Music Video Consumption on YouTube}.
\newblock \bibinfo{journal}{\emph{Social Science Computer Review}}
  \bibinfo{volume}{33}, \bibinfo{number}{6} (\bibinfo{year}{2015}),
  \bibinfo{pages}{730--748}.
\newblock
\showISSN{0894-4393}
\urldef\tempurl%
\url{https://doi.org/10.1177/0894439314562184}
\showDOI{\tempurl}


\bibitem[\protect\citeauthoryear{Batmaz, Yurekli, Bilge, and Kaleli}{Batmaz
  et~al\mbox{.}}{2019}]%
        {Batmaz2019_ai_review}
\bibfield{author}{\bibinfo{person}{Zeynep Batmaz}, \bibinfo{person}{Ali
  Yurekli}, \bibinfo{person}{Alper Bilge}, {and} \bibinfo{person}{Cihan
  Kaleli}.} \bibinfo{year}{2019}\natexlab{}.
\newblock \showarticletitle{A review on deep learning for recommender systems:
  challenges and remedies}.
\newblock \bibinfo{journal}{\emph{Artificial Intelligence Review}}
  \bibinfo{volume}{52}, \bibinfo{number}{1} (\bibinfo{date}{01 Jun}
  \bibinfo{year}{2019}), \bibinfo{pages}{1--37}.
\newblock
\showISSN{1573-7462}
\urldef\tempurl%
\url{https://doi.org/10.1007/s10462-018-9654-y}
\showDOI{\tempurl}


\bibitem[\protect\citeauthoryear{Bauer and Schedl}{Bauer and Schedl}{2018}]%
        {bauer_schedl_hicss2018}
\bibfield{author}{\bibinfo{person}{Christine Bauer} {and}
  \bibinfo{person}{Markus Schedl}.} \bibinfo{year}{2018}\natexlab{}.
\newblock \showarticletitle{On the importance of considering country-specific
  aspects on the online-market: an example of music recommendation considering
  country-specific mainstream}. In \bibinfo{booktitle}{\emph{51st Hawaii
  International Conference on System Sciences}} (Waikoloa, Big Island, HI, USA,
  3--6 January 2018). \bibinfo{pages}{3647--3656}.
\newblock
\showISBNx{978-0-9981331-1-9}
\urldef\tempurl%
\url{http://hdl.handle.net/10125/50349}
\showURL{%
\tempurl}
\newblock
\shownote{{http://hdl.handle.net/10125/50349}.}


\bibitem[\protect\citeauthoryear{Bauer and Schedl}{Bauer and Schedl}{2019}]%
        {10.1371/journal.pone.0217389}
\bibfield{author}{\bibinfo{person}{Christine Bauer} {and}
  \bibinfo{person}{Markus Schedl}.} \bibinfo{year}{2019}\natexlab{}.
\newblock \showarticletitle{Global and country-specific mainstreaminess
  measures: Definitions, analysis, and usage for improving personalized music
  recommendation systems}.
\newblock \bibinfo{journal}{\emph{PLOS ONE}} \bibinfo{volume}{14},
  \bibinfo{number}{6} (\bibinfo{date}{06} \bibinfo{year}{2019}),
  \bibinfo{pages}{1--36}.
\newblock
\urldef\tempurl%
\url{https://doi.org/10.1371/journal.pone.0217389}
\showDOI{\tempurl}


\bibitem[\protect\citeauthoryear{Bauer and Zangerle}{Bauer and
  Zangerle}{2019}]%
        {bauer_impactrs2019}
\bibfield{author}{\bibinfo{person}{Christine Bauer} {and} \bibinfo{person}{Eva
  Zangerle}.} \bibinfo{year}{2019}\natexlab{}.
\newblock \showarticletitle{Leveraging Multi-Method Evaluation for
  Multi-Stakeholder Settings}. In \bibinfo{booktitle}{\emph{1st Workshop on the
  Impact of Recommender Systems, co-located with 13th ACM Conference on
  Recommender Systems (ACM RecSys '19)}} (Copenhagen, Denmark, 19 September)
  \emph{(\bibinfo{series}{ImpactRS '19}, Vol.~\bibinfo{volume}{2462})},
  \bibfield{editor}{\bibinfo{person}{Oren~Sar Shalom}, \bibinfo{person}{Dietmar
  Jannach}, {and} \bibinfo{person}{Ido Guy}} (Eds.).
  \bibinfo{publisher}{Ceur-ws.org}.
\newblock
\urldef\tempurl%
\url{http://ceur-ws.org/Vol-2462/short3.pdf}
\showURL{%
\tempurl}
\newblock
\shownote{{http://ceur-ws.org/Vol-2462/short3.pdf}.}


\bibitem[\protect\citeauthoryear{Beer}{Beer}{2013}]%
        {Beer2013}
\bibfield{author}{\bibinfo{person}{David Beer}.}
  \bibinfo{year}{2013}\natexlab{}.
\newblock \showarticletitle{Genre, Boundary Drawing and the Classificatory
  Imagination}.
\newblock \bibinfo{journal}{\emph{Cultural Sociology}} \bibinfo{volume}{7},
  \bibinfo{number}{2} (\bibinfo{year}{2013}), \bibinfo{pages}{145--160}.
\newblock
\urldef\tempurl%
\url{https://doi.org/10.1177/1749975512473461}
\showDOI{\tempurl}
\showeprint{https://doi.org/10.1177/1749975512473461}


\bibitem[\protect\citeauthoryear{Bertin{-}Mahieux, Ellis, Whitman, and
  Lamere}{Bertin{-}Mahieux et~al\mbox{.}}{2011}]%
        {DBLP:conf/ismir/Bertin-MahieuxEWL11}
\bibfield{author}{\bibinfo{person}{Thierry Bertin{-}Mahieux},
  \bibinfo{person}{Daniel P.~W. Ellis}, \bibinfo{person}{Brian Whitman}, {and}
  \bibinfo{person}{Paul Lamere}.} \bibinfo{year}{2011}\natexlab{}.
\newblock \showarticletitle{The Million Song Dataset}. In
  \bibinfo{booktitle}{\emph{Proceedings of the 12th International Society for
  Music Information Retrieval Conference, {ISMIR} 2011, Miami, Florida, USA,
  October 24-28, 2011}}, \bibfield{editor}{\bibinfo{person}{Anssi Klapuri}
  {and} \bibinfo{person}{Colby Leider}} (Eds.). \bibinfo{publisher}{University
  of Miami}, \bibinfo{pages}{591--596}.
\newblock
\urldef\tempurl%
\url{http://ismir2011.ismir.net/papers/OS6-1.pdf}
\showURL{%
\tempurl}


\bibitem[\protect\citeauthoryear{Bonneville-Roussy, Rentfrow, Xu, and
  Potter}{Bonneville-Roussy et~al\mbox{.}}{2013}]%
        {BonnevilleRoussy2013}
\bibfield{author}{\bibinfo{person}{Arielle Bonneville-Roussy},
  \bibinfo{person}{P.~J. Rentfrow}, \bibinfo{person}{M.~K. Xu}, {and}
  \bibinfo{person}{J. Potter}.} \bibinfo{year}{2013}\natexlab{}.
\newblock \showarticletitle{Music through the ages: Trends in musical
  engagement and preferences from adolescence through middle adulthood}.
\newblock \bibinfo{journal}{\emph{Journal of Personality and Social
  Psychology}} \bibinfo{volume}{105}, \bibinfo{number}{4}
  (\bibinfo{year}{2013}), \bibinfo{pages}{703--717}.
\newblock
\showISSN{1939-1315 (Electronic), 0022-3514 (Linking)}
\urldef\tempurl%
\url{https://doi.org/10.1037/a0033770}
\showDOI{\tempurl}


\bibitem[\protect\citeauthoryear{Bonneville-Roussy and Rust}{Bonneville-Roussy
  and Rust}{2018}]%
        {BonnevilleRoussy2018_socialinfluences}
\bibfield{author}{\bibinfo{person}{Arielle Bonneville-Roussy} {and}
  \bibinfo{person}{John Rust}.} \bibinfo{year}{2018}\natexlab{}.
\newblock \showarticletitle{Age trends in musical preferences in adulthood: 2.
  Sources of social influences as determinants of preferences}.
\newblock \bibinfo{journal}{\emph{Musicae Scientiae}} \bibinfo{volume}{22},
  \bibinfo{number}{2} (\bibinfo{year}{2018}), \bibinfo{pages}{175--195}.
\newblock
\urldef\tempurl%
\url{https://doi.org/10.1177/1029864917704016}
\showDOI{\tempurl}


\bibitem[\protect\citeauthoryear{Brisson and Bianchi}{Brisson and
  Bianchi}{2019}]%
        {Brisson2019_genre}
\bibfield{author}{\bibinfo{person}{Romain Brisson} {and} \bibinfo{person}{Renzo
  Bianchi}.} \bibinfo{year}{2019}\natexlab{}.
\newblock \showarticletitle{On the relevance of music genre-based analysis in
  research on musical tastes}.
\newblock \bibinfo{journal}{\emph{Psychology of Music}} (\bibinfo{year}{2019}).
\newblock
\urldef\tempurl%
\url{https://doi.org/10.1177/0305735619828810}
\showDOI{\tempurl}
\showeprint{https://doi.org/10.1177/0305735619828810}


\bibitem[\protect\citeauthoryear{Bryson}{Bryson}{1997}]%
        {bryson1997univores}
\bibfield{author}{\bibinfo{person}{Bethany Bryson}.}
  \bibinfo{year}{1997}\natexlab{}.
\newblock \showarticletitle{What about the univores? Musical dislikes and
  group-based identity construction among Americans with low levels of
  education}.
\newblock \bibinfo{journal}{\emph{Poetics}} \bibinfo{volume}{25},
  \bibinfo{number}{2-3} (\bibinfo{year}{1997}), \bibinfo{pages}{141--156}.
\newblock


\bibitem[\protect\citeauthoryear{Budzinski and Pannicke}{Budzinski and
  Pannicke}{2017}]%
        {Budzinski2017}
\bibfield{author}{\bibinfo{person}{Oliver Budzinski} {and}
  \bibinfo{person}{Julia Pannicke}.} \bibinfo{year}{2017}\natexlab{}.
\newblock \showarticletitle{Do preferences for pop music converge across
  countries?: Empirical evidence from the Eurovision Song Contest}.
\newblock \bibinfo{journal}{\emph{Creative Industries Journal}}
  \bibinfo{volume}{10}, \bibinfo{number}{2} (\bibinfo{year}{2017}),
  \bibinfo{pages}{168--187}.
\newblock
\showISSN{1751-0694 1751-0708}
\urldef\tempurl%
\url{https://doi.org/10.1080/17510694.2017.1332451}
\showDOI{\tempurl}


\bibitem[\protect\citeauthoryear{{Central Intelligence Agency}}{{Central
  Intelligence Agency}}{2019}]%
        {cia_languages}
\bibfield{author}{\bibinfo{person}{{Central Intelligence Agency}}.}
  \bibinfo{year}{2019}\natexlab{}.
\newblock \showarticletitle{Languages}.
\newblock \bibinfo{journal}{\emph{The World Factbook}} (\bibinfo{year}{2019}).
\newblock
\newblock
\shownote{{https://www.cia.gov/library/publications/resources/the-world-factbook/fields/402.html}.}


\bibitem[\protect\citeauthoryear{Cheng, Shen, Nie, Chua, and Kankanhalli}{Cheng
  et~al\mbox{.}}{2017}]%
        {Cheng:2017:EUI:3077136.3080772}
\bibfield{author}{\bibinfo{person}{Zhiyong Cheng}, \bibinfo{person}{Jialie
  Shen}, \bibinfo{person}{Liqiang Nie}, \bibinfo{person}{Tat-Seng Chua}, {and}
  \bibinfo{person}{Mohan Kankanhalli}.} \bibinfo{year}{2017}\natexlab{}.
\newblock \showarticletitle{Exploring User-Specific Information in Music
  Retrieval}. In \bibinfo{booktitle}{\emph{40th International ACM SIGIR
  Conference on Research and Development in Information Retrieval}} (Shinjuku,
  Tokyo, Japan) \emph{(\bibinfo{series}{SIGIR '17})}. \bibinfo{publisher}{ACM},
  \bibinfo{address}{New York, NY, USA}, \bibinfo{pages}{655--664}.
\newblock
\showISBNx{978-1-4503-5022-8}
\urldef\tempurl%
\url{https://doi.org/10.1145/3077136.3080772}
\showDOI{\tempurl}


\bibitem[\protect\citeauthoryear{Colley}{Colley}{2008}]%
        {colley2008}
\bibfield{author}{\bibinfo{person}{A. Colley}.}
  \bibinfo{year}{2008}\natexlab{}.
\newblock \showarticletitle{Young people's musical taste: Relationship with
  gender and gender-related traits}.
\newblock \bibinfo{journal}{\emph{Journal of Applied Social Psychology}}
  \bibinfo{volume}{38}, \bibinfo{number}{8} (\bibinfo{year}{2008}),
  \bibinfo{pages}{2039--2055}.
\newblock
\showISSN{0021-9029}
\urldef\tempurl%
\url{https://doi.org/10.1111/j.1559-1816.2008.00379.x}
\showDOI{\tempurl}


\bibitem[\protect\citeauthoryear{Coulangeon}{Coulangeon}{2003}]%
        {Coulangeon2003}
\bibfield{author}{\bibinfo{person}{Philippe Coulangeon}.}
  \bibinfo{year}{2003}\natexlab{}.
\newblock \showarticletitle{La stratification sociale des go{\^u}ts musicaux.
  Le mod{\`e}le de la l{\'e}gitimit{\'e}culturelle en question}.
\newblock \bibinfo{journal}{\emph{Revue Fran{\c c}aise de Sociologie}}
  \bibinfo{volume}{44}, \bibinfo{number}{1} (\bibinfo{year}{2003}),
  \bibinfo{pages}{3--33}.
\newblock
\urldef\tempurl%
\url{https://doi.org/10.3917/rfs.441.0003}
\showDOI{\tempurl}


\bibitem[\protect\citeauthoryear{Coulangeon}{Coulangeon}{2005}]%
        {Coulangeon2005}
\bibfield{author}{\bibinfo{person}{Philippe Coulangeon}.}
  \bibinfo{year}{2005}\natexlab{}.
\newblock \showarticletitle{Social Stratification of Musical Tastes:
  Questioning the Cultural Legitimacy Model}.
\newblock \bibinfo{journal}{\emph{Revue Fran{\c c}aise de Sociologie}}
  \bibinfo{volume}{46} (\bibinfo{date}{5} \bibinfo{year}{2005}),
  \bibinfo{pages}{123--154}.
\newblock
\urldef\tempurl%
\url{https://doi.org/10.3917/rfs.465.0123}
\showDOI{\tempurl}


\bibitem[\protect\citeauthoryear{Coulangeon and Roharik}{Coulangeon and
  Roharik}{2005}]%
        {coulangeon:hal-01053502}
\bibfield{author}{\bibinfo{person}{Philippe Coulangeon} {and}
  \bibinfo{person}{Ionela Roharik}.} \bibinfo{year}{2005}\natexlab{}.
\newblock \showarticletitle{Testing the ``Omnivore/Univore'' Hypothesis in a
  Cross-National Perspective. On the Social Meaning of Ecletism in Musical
  Tastes}. In \bibinfo{booktitle}{\emph{{The Summer Meeting of the ISA RC28,
  UCLA}}}.
\newblock
\urldef\tempurl%
\url{https://hal-sciencespo.archives-ouvertes.fr/hal-01053502}
\showURL{%
\tempurl}


\bibitem[\protect\citeauthoryear{Cunningham, Downie, and Bainbridge}{Cunningham
  et~al\mbox{.}}{2005}]%
        {cunningham_etal:ismir:2005}
\bibfield{author}{\bibinfo{person}{Sally~Jo Cunningham},
  \bibinfo{person}{J.~Stephen Downie}, {and} \bibinfo{person}{David
  Bainbridge}.} \bibinfo{year}{2005}\natexlab{}.
\newblock \showarticletitle{{``The Pain, The Pain": Modelling Music Information
  Behavior And The Songs We Hate}}. In \bibinfo{booktitle}{\emph{Proceedings of
  the 6th International Conference on Music Information Retrieval {(ISMIR
  2005)}}}. \bibinfo{address}{London, UK}, \bibinfo{pages}{474--477}.
\newblock


\bibitem[\protect\citeauthoryear{Dacrema, Cremonesi, and Jannach}{Dacrema
  et~al\mbox{.}}{2019}]%
        {10.1145/3298689.3347058}
\bibfield{author}{\bibinfo{person}{Maurizio~Ferrari Dacrema},
  \bibinfo{person}{Paolo Cremonesi}, {and} \bibinfo{person}{Dietmar Jannach}.}
  \bibinfo{year}{2019}\natexlab{}.
\newblock \showarticletitle{{Are We Really Making Much Progress? A Worrying
  Analysis of Recent Neural Recommendation Approaches}}. In
  \bibinfo{booktitle}{\emph{Proceedings of the 13th ACM Conference on
  Recommender Systems}} (Copenhagen, Denmark) \emph{(\bibinfo{series}{RecSys
  ’19})}. \bibinfo{publisher}{Association for Computing Machinery},
  \bibinfo{address}{New York, NY, USA}, \bibinfo{pages}{101–109}.
\newblock
\showISBNx{9781450362436}
\urldef\tempurl%
\url{https://doi.org/10.1145/3298689.3347058}
\showDOI{\tempurl}


\bibitem[\protect\citeauthoryear{DiMaggio}{DiMaggio}{1982}]%
        {Dimaggio1982}
\bibfield{author}{\bibinfo{person}{Paul DiMaggio}.}
  \bibinfo{year}{1982}\natexlab{}.
\newblock \showarticletitle{Cultural entrepreneurship in nineteenth-century
  Boston: the creation of an organizational base for high culture in America}.
\newblock \bibinfo{journal}{\emph{Media, Culture \& Society}}
  \bibinfo{volume}{4}, \bibinfo{number}{1} (\bibinfo{year}{1982}),
  \bibinfo{pages}{33--50}.
\newblock
\urldef\tempurl%
\url{https://doi.org/10.1177/016344378200400104}
\showDOI{\tempurl}
\showeprint{https://doi.org/10.1177/016344378200400104}


\bibitem[\protect\citeauthoryear{Dolata}{Dolata}{2013}]%
        {dolata2013transformative}
\bibfield{author}{\bibinfo{person}{Ulrich Dolata}.}
  \bibinfo{year}{2013}\natexlab{}.
\newblock \bibinfo{booktitle}{\emph{The transformative capacity of new
  technologies: A theory of sociotechnical change}}. \bibinfo{series}{Routledge
  Advances in Sociology}, Vol.~\bibinfo{volume}{96}.
\newblock \bibinfo{publisher}{Routledge}, \bibinfo{address}{London, United
  Kingdom}.
\newblock


\bibitem[\protect\citeauthoryear{Fisher and Preece}{Fisher and Preece}{2003}]%
        {FISHER200369}
\bibfield{author}{\bibinfo{person}{Timothy~C.G. Fisher} {and}
  \bibinfo{person}{Stephen~B. Preece}.} \bibinfo{year}{2003}\natexlab{}.
\newblock \showarticletitle{Evolution, extinction, or status quo? Canadian
  performing arts audiences in the 1990s}.
\newblock \bibinfo{journal}{\emph{Poetics}} \bibinfo{volume}{31},
  \bibinfo{number}{2} (\bibinfo{year}{2003}), \bibinfo{pages}{69--86}.
\newblock
\showISSN{0304-422X}
\urldef\tempurl%
\url{https://doi.org/10.1016/S0304-422X(03)00004-4}
\showDOI{\tempurl}


\bibitem[\protect\citeauthoryear{Friedman}{Friedman}{1937}]%
        {Friedman1937}
\bibfield{author}{\bibinfo{person}{Milton Friedman}.}
  \bibinfo{year}{1937}\natexlab{}.
\newblock \showarticletitle{The Use of Ranks to Avoid the Assumption of
  Normality Implicit in the Analysis of Variance}.
\newblock \bibinfo{journal}{\emph{J. Amer. Statist. Assoc.}}
  \bibinfo{volume}{32}, \bibinfo{number}{200} (\bibinfo{year}{1937}),
  \bibinfo{pages}{675--701}.
\newblock
\urldef\tempurl%
\url{https://doi.org/10.1080/01621459.1937.10503522}
\showDOI{\tempurl}
\showeprint{https://www.tandfonline.com/doi/pdf/10.1080/01621459.1937.10503522}


\bibitem[\protect\citeauthoryear{Gomez~Herrera and Martens}{Gomez~Herrera and
  Martens}{2018}]%
        {gomez2018_copyright}
\bibfield{author}{\bibinfo{person}{Estrella Gomez~Herrera} {and}
  \bibinfo{person}{Bertin Martens}.} \bibinfo{year}{2018}\natexlab{}.
\newblock \showarticletitle{Language, Copyright and Geographic Segmentation in
  the EU Digital, Single Market for Music and Film}.
\newblock \bibinfo{journal}{\emph{Review of Economic Research on Copyright
  Issues}} \bibinfo{volume}{15}, \bibinfo{number}{1} (\bibinfo{year}{2018}),
  \bibinfo{pages}{20--37}.
\newblock
\urldef\tempurl%
\url{https://ssrn.com/abstract=3243751}
\showURL{%
\tempurl}


\bibitem[\protect\citeauthoryear{Halko, Martinsson, and Tropp}{Halko
  et~al\mbox{.}}{201}]%
        {halko2011finding}
\bibfield{author}{\bibinfo{person}{Nathan Halko}, \bibinfo{person}{Per-Gunnar
  Martinsson}, {and} \bibinfo{person}{Joel~A. Tropp}.}
  \bibinfo{year}{201}\natexlab{}.
\newblock \showarticletitle{Finding structure with randomness: Probabilistic
  algorithms for constructing approximate matrix decompositions}.
\newblock \bibinfo{journal}{\emph{SIAM Rev.}}  \bibinfo{volume}{53}
  (\bibinfo{year}{201}), \bibinfo{pages}{217--288}.
\newblock
\urldef\tempurl%
\url{https://doi.org/10.1137/090771806}
\showDOI{\tempurl}


\bibitem[\protect\citeauthoryear{Helliwell, Layard, and Sachs}{Helliwell
  et~al\mbox{.}}{2016}]%
        {helliwell2016world}
\bibfield{author}{\bibinfo{person}{John~F Helliwell}, \bibinfo{person}{Peter~RG
  Layard}, {and} \bibinfo{person}{Jeffrey Sachs}.}
  \bibinfo{year}{2016}\natexlab{}.
\newblock \bibinfo{booktitle}{\emph{World happiness report 2016 update}}.
\newblock \bibinfo{publisher}{Sustainable Development Solutions Network}.
\newblock


\bibitem[\protect\citeauthoryear{Hofstede, Hofstede, and Minkov}{Hofstede
  et~al\mbox{.}}{2005}]%
        {hofstede2005cultures}
\bibfield{author}{\bibinfo{person}{Geert Hofstede}, \bibinfo{person}{Gert~Jan
  Hofstede}, {and} \bibinfo{person}{Michael Minkov}.}
  \bibinfo{year}{2005}\natexlab{}.
\newblock \bibinfo{booktitle}{\emph{Cultures and organizations: Software of the
  mind}}. Vol.~\bibinfo{volume}{2}.
\newblock \bibinfo{publisher}{McGraw-Hill}, \bibinfo{address}{New York, NY
  USA}.
\newblock


\bibitem[\protect\citeauthoryear{Holbrook, Weiss, and Habich}{Holbrook
  et~al\mbox{.}}{2002}]%
        {Holbrook2002}
\bibfield{author}{\bibinfo{person}{Morris~B. Holbrook},
  \bibinfo{person}{Michael~J. Weiss}, {and} \bibinfo{person}{John Habich}.}
  \bibinfo{year}{2002}\natexlab{}.
\newblock \showarticletitle{Disentangling Effacement, Omnivore, and Distinction
  Effects on the Consumption of Cultural Activities: An Illustration}.
\newblock \bibinfo{journal}{\emph{Marketing Letters}} \bibinfo{volume}{13},
  \bibinfo{number}{4} (\bibinfo{year}{2002}), \bibinfo{pages}{345--357}.
\newblock
\showISBNx{1573-059X}
\urldef\tempurl%
\url{https://doi.org/10.1023/A:1020322600709}
\showDOI{\tempurl}


\bibitem[\protect\citeauthoryear{Horkheimer and Adorno}{Horkheimer and
  Adorno}{1972}]%
        {horkheimer1972dialectic}
\bibfield{author}{\bibinfo{person}{Max Horkheimer} {and}
  \bibinfo{person}{Theodor~W Adorno}.} \bibinfo{year}{1972}\natexlab{}.
\newblock \bibinfo{booktitle}{\emph{Dialectic of Enlightenment}}.
\newblock \bibinfo{publisher}{Seabury Press New York}, \bibinfo{address}{New
  York, NY, USA}.
\newblock


\bibitem[\protect\citeauthoryear{Hracs, Seman, and Virani}{Hracs
  et~al\mbox{.}}{2017}]%
        {hracs2017_musicproduction_consumpation}
\bibfield{author}{\bibinfo{person}{Brian~J. Hracs}, \bibinfo{person}{Michael
  Seman}, {and} \bibinfo{person}{Tarek~E. Virani}.}
  \bibinfo{year}{2017}\natexlab{}.
\newblock \bibinfo{booktitle}{\emph{The Production and Consumption of Music in
  the Digital Age}}.
\newblock \bibinfo{publisher}{Routledge}, \bibinfo{address}{New York, NY, USA}.
\newblock


\bibitem[\protect\citeauthoryear{Johnston}{Johnston}{2018}]%
        {spotify_microgenres2018}
\bibfield{author}{\bibinfo{person}{Maura Johnston}.}
  \bibinfo{year}{2018}\natexlab{}.
\newblock \showarticletitle{{How Spotify Discovers the Genres of Tomorrow}}.
\newblock  (\bibinfo{year}{2018}).
\newblock
\newblock
\shownote{{https://artists.spotify.com/blog/how-spotify-discovers-the-genres-of-tomorrow}.}


\bibitem[\protect\citeauthoryear{Jones}{Jones}{1972}]%
        {Jones72astatistical}
\bibfield{author}{\bibinfo{person}{Karen~Sp\"{a}rck Jones}.}
  \bibinfo{year}{1972}\natexlab{}.
\newblock \showarticletitle{{A statistical interpretation of term specificity
  and its application in retrieval}}.
\newblock \bibinfo{journal}{\emph{Journal of Documentation}}
  \bibinfo{volume}{28} (\bibinfo{year}{1972}), \bibinfo{pages}{11--21}.
\newblock


\bibitem[\protect\citeauthoryear{Jordan, Ghahramani, Jaakkola, and Saul}{Jordan
  et~al\mbox{.}}{1999}]%
        {jordan1999introduction}
\bibfield{author}{\bibinfo{person}{Michael~I Jordan}, \bibinfo{person}{Zoubin
  Ghahramani}, \bibinfo{person}{Tommi~S Jaakkola}, {and}
  \bibinfo{person}{Lawrence~K Saul}.} \bibinfo{year}{1999}\natexlab{}.
\newblock \showarticletitle{An introduction to variational methods for
  graphical models}.
\newblock \bibinfo{journal}{\emph{Machine learning}} \bibinfo{volume}{37},
  \bibinfo{number}{2} (\bibinfo{year}{1999}), \bibinfo{pages}{183--233}.
\newblock


\bibitem[\protect\citeauthoryear{Katz-Gerro}{Katz-Gerro}{1999}]%
        {Katz-Gerro1999}
\bibfield{author}{\bibinfo{person}{Tally Katz-Gerro}.}
  \bibinfo{year}{1999}\natexlab{}.
\newblock \showarticletitle{Cultural Consumption and Social Stratification:
  Leisure Activities, Musical Tastes, and Social Location}.
\newblock \bibinfo{journal}{\emph{Sociological Perspectives}}
  \bibinfo{volume}{42}, \bibinfo{number}{4} (\bibinfo{year}{1999}),
  \bibinfo{pages}{627--646}.
\newblock
\urldef\tempurl%
\url{https://doi.org/10.2307/1389577}
\showDOI{\tempurl}
\showeprint{https://doi.org/10.2307/1389577}


\bibitem[\protect\citeauthoryear{Katz-Gerro}{Katz-Gerro}{2002}]%
        {Katz-Gerro2002}
\bibfield{author}{\bibinfo{person}{Tally Katz-Gerro}.}
  \bibinfo{year}{2002}\natexlab{}.
\newblock \showarticletitle{Highbrow Cultural Consumption and Class Distinction
  in Italy, Israel, West Germany, Sweden, and the United States}.
\newblock \bibinfo{journal}{\emph{Social Forces}} \bibinfo{volume}{81},
  \bibinfo{number}{1} (\bibinfo{date}{09} \bibinfo{year}{2002}),
  \bibinfo{pages}{207--229}.
\newblock
\showISSN{0037-7732}
\urldef\tempurl%
\url{https://doi.org/10.1353/sof.2002.0050}
\showDOI{\tempurl}
\showeprint{https://academic.oup.com/sf/article-pdf/81/1/207/6519885/81-1-207.pdf}


\bibitem[\protect\citeauthoryear{Koren, Bell, and Volinsky}{Koren
  et~al\mbox{.}}{2009}]%
        {Koren:2009:MFT:1608565.1608614}
\bibfield{author}{\bibinfo{person}{Yehuda Koren}, \bibinfo{person}{Robert
  Bell}, {and} \bibinfo{person}{Chris Volinsky}.}
  \bibinfo{year}{2009}\natexlab{}.
\newblock \showarticletitle{{Matrix Factorization Techniques for Recommender
  Systems}}.
\newblock \bibinfo{journal}{\emph{Computer}} \bibinfo{volume}{42},
  \bibinfo{number}{8} (\bibinfo{date}{Aug.} \bibinfo{year}{2009}),
  \bibinfo{pages}{30--37}.
\newblock
\showISSN{0018-9162}
\urldef\tempurl%
\url{https://doi.org/10.1109/MC.2009.263}
\showDOI{\tempurl}


\bibitem[\protect\citeauthoryear{Kramer}{Kramer}{1991}]%
        {kramer1991nonlinear}
\bibfield{author}{\bibinfo{person}{Mark~A Kramer}.}
  \bibinfo{year}{1991}\natexlab{}.
\newblock \showarticletitle{Nonlinear principal component analysis using
  autoassociative neural networks}.
\newblock \bibinfo{journal}{\emph{AIChE journal}} \bibinfo{volume}{37},
  \bibinfo{number}{2} (\bibinfo{year}{1991}), \bibinfo{pages}{233--243}.
\newblock


\bibitem[\protect\citeauthoryear{Lai, Lee, and Huang}{Lai
  et~al\mbox{.}}{2019}]%
        {DBLP:journals/ijmms/LaiLH19}
\bibfield{author}{\bibinfo{person}{Chin{-}Hui Lai}, \bibinfo{person}{Shin{-}Jye
  Lee}, {and} \bibinfo{person}{Hung{-}Ling Huang}.}
  \bibinfo{year}{2019}\natexlab{}.
\newblock \showarticletitle{A social recommendation method based on the
  integration of social relationship and product popularity}.
\newblock \bibinfo{journal}{\emph{International Journal of Human-Computer
  Studies}}  \bibinfo{volume}{121} (\bibinfo{year}{2019}),
  \bibinfo{pages}{42--57}.
\newblock
\urldef\tempurl%
\url{https://doi.org/10.1016/j.ijhcs.2018.04.002}
\showDOI{\tempurl}


\bibitem[\protect\citeauthoryear{Lee and Hu}{Lee and Hu}{2014}]%
        {jinhalee2014}
\bibfield{author}{\bibinfo{person}{Jin~Ha Lee} {and} \bibinfo{person}{Xiao
  Hu}.} \bibinfo{year}{2014}\natexlab{}.
\newblock \showarticletitle{Cross-cultural Similarities and Differences in
  Music Mood Perception}. In \bibinfo{booktitle}{\emph{iConference 2014}}.
  \bibinfo{publisher}{iSchools}, \bibinfo{pages}{249--269}.
\newblock
\urldef\tempurl%
\url{https://doi.org/10.9776/14081}
\showDOI{\tempurl}


\bibitem[\protect\citeauthoryear{Levine}{Levine}{1988}]%
        {levine1988highbrow}
\bibfield{author}{\bibinfo{person}{Lawrence~W Levine}.}
  \bibinfo{year}{1988}\natexlab{}.
\newblock \bibinfo{booktitle}{\emph{Highbrow/lowbrow: the emergence of cultural
  hierarchy in America}}.
\newblock \bibinfo{publisher}{Harvard University Press},
  \bibinfo{address}{Cambridge, MA, USA}.
\newblock


\bibitem[\protect\citeauthoryear{Liang, Krishnan, Hoffman, and Jebara}{Liang
  et~al\mbox{.}}{2018}]%
        {Liang:2018:VAC:3178876.3186150}
\bibfield{author}{\bibinfo{person}{Dawen Liang}, \bibinfo{person}{Rahul~G.
  Krishnan}, \bibinfo{person}{Matthew~D. Hoffman}, {and} \bibinfo{person}{Tony
  Jebara}.} \bibinfo{year}{2018}\natexlab{}.
\newblock \showarticletitle{Variational Autoencoders for Collaborative
  Filtering}. In \bibinfo{booktitle}{\emph{Proceedings of the 2018 World Wide
  Web Conference}} (Lyon, France) \emph{(\bibinfo{series}{WWW '18})}.
  \bibinfo{publisher}{International World Wide Web Conferences Steering
  Committee}, \bibinfo{address}{Republic and Canton of Geneva, Switzerland},
  \bibinfo{pages}{689--698}.
\newblock
\showISBNx{978-1-4503-5639-8}
\urldef\tempurl%
\url{https://doi.org/10.1145/3178876.3186150}
\showDOI{\tempurl}


\bibitem[\protect\citeauthoryear{L{\'o}pez-Sintas, Cebollada, Filimon, and
  Gharhaman}{L{\'o}pez-Sintas et~al\mbox{.}}{2014}]%
        {LOPEZSINTAS201456}
\bibfield{author}{\bibinfo{person}{Jordi L{\'o}pez-Sintas},
  \bibinfo{person}{{\`A}ngel Cebollada}, \bibinfo{person}{Nela Filimon}, {and}
  \bibinfo{person}{Abaghan Gharhaman}.} \bibinfo{year}{2014}\natexlab{}.
\newblock \showarticletitle{Music access patterns: A social interpretation}.
\newblock \bibinfo{journal}{\emph{Poetics}}  \bibinfo{volume}{46}
  (\bibinfo{year}{2014}), \bibinfo{pages}{56--74}.
\newblock
\showISSN{0304-422X}
\urldef\tempurl%
\url{https://doi.org/10.1016/j.poetic.2014.09.003}
\showDOI{\tempurl}


\bibitem[\protect\citeauthoryear{Moore, Joachims, and Turnbull}{Moore
  et~al\mbox{.}}{2014}]%
        {moore_etal:ismir:2014}
\bibfield{author}{\bibinfo{person}{Joshua~L. Moore}, \bibinfo{person}{Thorsten
  Joachims}, {and} \bibinfo{person}{Douglas Turnbull}.}
  \bibinfo{year}{2014}\natexlab{}.
\newblock \showarticletitle{{Taste Space Versus the World: An Embedding
  Analysis of Listening Habits and Geography}}. In
  \bibinfo{booktitle}{\emph{Proceedings of the 15th International Society for
  Music Information Retrieval Conference (ISMIR 2014)}}.
  \bibinfo{address}{Taipei, Taiwan}.
\newblock


\bibitem[\protect\citeauthoryear{Morrison and Demorest}{Morrison and
  Demorest}{2009}]%
        {Morrison2009}
\bibfield{author}{\bibinfo{person}{Steven~J. Morrison} {and}
  \bibinfo{person}{Steven~M. Demorest}.} \bibinfo{year}{2009}\natexlab{}.
\newblock \showarticletitle{Cultural constraints on music perception and
  cognition}.
\newblock \bibinfo{journal}{\emph{Progress in Brain Research}}
  \bibinfo{volume}{178} (\bibinfo{year}{2009}), \bibinfo{pages}{67--77}.
\newblock
\urldef\tempurl%
\url{https://doi.org/10.1016/S0079-6123(09)17805-6}
\showDOI{\tempurl}


\bibitem[\protect\citeauthoryear{Nuccio, Guerzoni, and Katz-Gerro}{Nuccio
  et~al\mbox{.}}{2018}]%
        {Nuccio2018}
\bibfield{author}{\bibinfo{person}{Massimiliano Nuccio}, \bibinfo{person}{Marco
  Guerzoni}, {and} \bibinfo{person}{Tally Katz-Gerro}.}
  \bibinfo{year}{2018}\natexlab{}.
\newblock \showarticletitle{Beyond Class Stratification: The Rise of the
  Eclectic Music Consumer in the Modern Age}.
\newblock \bibinfo{journal}{\emph{Cultural Sociology}} \bibinfo{volume}{12},
  \bibinfo{number}{3} (\bibinfo{year}{2018}), \bibinfo{pages}{343--367}.
\newblock
\urldef\tempurl%
\url{https://doi.org/10.1177/1749975518786039}
\showDOI{\tempurl}
\showeprint{https://doi.org/10.1177/1749975518786039}


\bibitem[\protect\citeauthoryear{Peterson and Kern}{Peterson and Kern}{1996}]%
        {peterson_kern_1996}
\bibfield{author}{\bibinfo{person}{Richard~A. Peterson} {and}
  \bibinfo{person}{Roger~M. Kern}.} \bibinfo{year}{1996}\natexlab{}.
\newblock \showarticletitle{Changing Highbrow Taste: From Snob to Omnivore}.
\newblock \bibinfo{journal}{\emph{American Sociological Review}}
  \bibinfo{volume}{61}, \bibinfo{number}{5} (\bibinfo{year}{1996}),
  \bibinfo{pages}{900--907}.
\newblock
\showISSN{00031224}
\urldef\tempurl%
\url{http://www.jstor.org/stable/2096460}
\showURL{%
\tempurl}


\bibitem[\protect\citeauthoryear{Peterson and Simkus}{Peterson and
  Simkus}{1992}]%
        {peterson1992seven}
\bibfield{author}{\bibinfo{person}{Richard~A Peterson} {and}
  \bibinfo{person}{Albert Simkus}.} \bibinfo{year}{1992}\natexlab{}.
\newblock \showarticletitle{How musical tastes mark occupational status
  groups}.
\newblock In \bibinfo{booktitle}{\emph{Cultivating differences: Symbolic
  boundaries and the making of inequality}}. Vol.~\bibinfo{volume}{152}.
  \bibinfo{publisher}{University of Chicago Press}, \bibinfo{address}{Chicago,
  IL, USA}, Chapter~7, \bibinfo{pages}{152--186}.
\newblock


\bibitem[\protect\citeauthoryear{{Pichl}, {Zangerle}, {Specht}, and
  {Schedl}}{{Pichl} et~al\mbox{.}}{2017}]%
        {pichl2017_ism}
\bibfield{author}{\bibinfo{person}{Michael {Pichl}}, \bibinfo{person}{Eva
  {Zangerle}}, \bibinfo{person}{G{\"u}nther {Specht}}, {and}
  \bibinfo{person}{Markus {Schedl}}.} \bibinfo{year}{2017}\natexlab{}.
\newblock \showarticletitle{Mining Culture-Specific Music Listening Behavior
  from Social Media Data}. In \bibinfo{booktitle}{\emph{2017 IEEE International
  Symposium on Multimedia (ISM)}} \emph{(\bibinfo{series}{ISM 2017})}. IEEE,
  \bibinfo{pages}{208--215}.
\newblock
\urldef\tempurl%
\url{https://doi.org/10.1109/ISM.2017.35}
\showDOI{\tempurl}


\bibitem[\protect\citeauthoryear{Rentfrow and Gosling}{Rentfrow and
  Gosling}{2003}]%
        {Rentfrow_doremi}
\bibfield{author}{\bibinfo{person}{P.~J. Rentfrow} {and} \bibinfo{person}{S.~D.
  Gosling}.} \bibinfo{year}{2003}\natexlab{}.
\newblock \showarticletitle{The do re mi's of everyday life: the structure and
  personality correlates of music preferences}.
\newblock \bibinfo{journal}{\emph{Journal of Personality and Social
  Psychology}} \bibinfo{volume}{84}, \bibinfo{number}{6}
  (\bibinfo{year}{2003}), \bibinfo{pages}{1236--1256}.
\newblock
\showISSN{0022-3514}
\urldef\tempurl%
\url{https://doi.org/10.1037/0022-3514.84.6.1236}
\showDOI{\tempurl}


\bibitem[\protect\citeauthoryear{Ricci, Rokach, Shapira, and Kantor}{Ricci
  et~al\mbox{.}}{2015}]%
        {ricci_etal:rsh:2015}
\bibfield{editor}{\bibinfo{person}{Francesco Ricci}, \bibinfo{person}{Lior
  Rokach}, \bibinfo{person}{Bracha Shapira}, {and} \bibinfo{person}{Paul~B.
  Kantor}} (Eds.). \bibinfo{year}{2015}\natexlab{}.
\newblock \bibinfo{booktitle}{\emph{{Recommender Systems Handbook}}
  (\bibinfo{edition}{2nd} ed.)}.
\newblock \bibinfo{publisher}{Springer}.
\newblock


\bibitem[\protect\citeauthoryear{Rossman and Peterson}{Rossman and
  Peterson}{2015}]%
        {ROSSMAN2015139}
\bibfield{author}{\bibinfo{person}{Gabriel Rossman} {and}
  \bibinfo{person}{Richard~A. Peterson}.} \bibinfo{year}{2015}\natexlab{}.
\newblock \showarticletitle{The instability of omnivorous cultural taste over
  time}.
\newblock \bibinfo{journal}{\emph{Poetics}}  \bibinfo{volume}{52}
  (\bibinfo{year}{2015}), \bibinfo{pages}{139--153}.
\newblock
\showISSN{0304-422X}
\urldef\tempurl%
\url{https://doi.org/10.1016/j.poetic.2015.05.004}
\showDOI{\tempurl}


\bibitem[\protect\citeauthoryear{S{\'a}nchez-Moreno, Gonz{\'a}lez, Vicente,
  Batista, and Garc{\'\i}a}{S{\'a}nchez-Moreno et~al\mbox{.}}{2016}]%
        {SANCHEZMORENO2016234}
\bibfield{author}{\bibinfo{person}{Diego S{\'a}nchez-Moreno},
  \bibinfo{person}{Ana B.~Gil Gonz{\'a}lez},
  \bibinfo{person}{M.~Dolores~Mu{\~n}oz Vicente}, \bibinfo{person}{Vivian
  F.~L{\'o}pez Batista}, {and} \bibinfo{person}{Mar{\'\i}a N.~Moreno
  Garc{\'\i}a}.} \bibinfo{year}{2016}\natexlab{}.
\newblock \showarticletitle{A collaborative filtering method for music
  recommendation using playing coefficients for artists and users}.
\newblock \bibinfo{journal}{\emph{Expert Systems with Applications}}
  \bibinfo{volume}{66} (\bibinfo{year}{2016}), \bibinfo{pages}{234--244}.
\newblock
\showISSN{0957-4174}
\urldef\tempurl%
\url{https://doi.org/10.1016/j.eswa.2016.09.019}
\showDOI{\tempurl}


\bibitem[\protect\citeauthoryear{Sch{\"a}fer and Mehlhorn}{Sch{\"a}fer and
  Mehlhorn}{2017}]%
        {schafer2017can}
\bibfield{author}{\bibinfo{person}{Thomas Sch{\"a}fer} {and}
  \bibinfo{person}{Claudia Mehlhorn}.} \bibinfo{year}{2017}\natexlab{}.
\newblock \showarticletitle{Can personality traits predict musical style
  preferences? A meta-analysis}.
\newblock \bibinfo{journal}{\emph{Personality and Individual Differences}}
  \bibinfo{volume}{116} (\bibinfo{year}{2017}), \bibinfo{pages}{265--273}.
\newblock
\urldef\tempurl%
\url{https://doi.org/10.1016/j.paid.2017.04.061}
\showDOI{\tempurl}


\bibitem[\protect\citeauthoryear{Schedl}{Schedl}{2016}]%
        {Schedl:2016:LDM:2911996.2912004}
\bibfield{author}{\bibinfo{person}{Markus Schedl}.}
  \bibinfo{year}{2016}\natexlab{}.
\newblock \showarticletitle{{The LFM-1B Dataset for Music Retrieval and
  Recommendation}}. In \bibinfo{booktitle}{\emph{Proceedings of the 2016 ACM on
  International Conference on Multimedia Retrieval}} (New York, New York, USA)
  \emph{(\bibinfo{series}{ICMR '16})}. \bibinfo{publisher}{ACM},
  \bibinfo{address}{New York, NY, USA}, \bibinfo{pages}{103--110}.
\newblock
\showISBNx{978-1-4503-4359-6}
\urldef\tempurl%
\url{https://doi.org/10.1145/2911996.2912004}
\showDOI{\tempurl}


\bibitem[\protect\citeauthoryear{Schedl}{Schedl}{2017}]%
        {Schedl2017}
\bibfield{author}{\bibinfo{person}{Markus Schedl}.}
  \bibinfo{year}{2017}\natexlab{}.
\newblock \showarticletitle{Investigating country-specific music preferences
  and music recommendation algorithms with the LFM-1b dataset}.
\newblock \bibinfo{journal}{\emph{International Journal of Multimedia
  Information Retrieval}} \bibinfo{volume}{6}, \bibinfo{number}{1}
  (\bibinfo{date}{March} \bibinfo{year}{2017}), \bibinfo{pages}{71--84}.
\newblock
\showISSN{2192-662X}
\urldef\tempurl%
\url{https://doi.org/10.1007/s13735-017-0118-y}
\showDOI{\tempurl}


\bibitem[\protect\citeauthoryear{Schedl}{Schedl}{2019}]%
        {10.3389/fams.2019.00044}
\bibfield{author}{\bibinfo{person}{Markus Schedl}.}
  \bibinfo{year}{2019}\natexlab{}.
\newblock \showarticletitle{{Deep Learning in Music Recommendation Systems}}.
\newblock \bibinfo{journal}{\emph{Frontiers in Applied Mathematics and
  Statistics}}  \bibinfo{volume}{5} (\bibinfo{year}{2019}),
  \bibinfo{pages}{44}.
\newblock
\showISSN{2297-4687}
\urldef\tempurl%
\url{https://doi.org/10.3389/fams.2019.00044}
\showDOI{\tempurl}


\bibitem[\protect\citeauthoryear{Schedl and Bauer}{Schedl and Bauer}{2017}]%
        {schedl2017_kidrec}
\bibfield{author}{\bibinfo{person}{Markus Schedl} {and}
  \bibinfo{person}{Christine Bauer}.} \bibinfo{year}{2017}\natexlab{}.
\newblock \showarticletitle{Online Music Listening Culture of Kids and
  Adolescents: Listening Analysis and Music Recommendation Tailored to the
  Young}. In \bibinfo{booktitle}{\emph{1st International Workshop on Children
  and Recommender Systems, in conjunction with 11th ACM Conference on
  Recommender Systems (RecSys '17)}} (Como, Italy, 27 August)
  \emph{(\bibinfo{series}{KidRec '17})}.
\newblock
\showeprint[arxiv]{1912.11564}~[cs.IR]
\urldef\tempurl%
\url{https://arxiv.org/abs/1912.11564}
\showURL{%
\tempurl}


\bibitem[\protect\citeauthoryear{Schedl, Bauer, Reisinger, Kowald, and
  Lex}{Schedl et~al\mbox{.}}{2020}]%
        {schedl_markus_2020_3832071}
\bibfield{author}{\bibinfo{person}{Markus Schedl}, \bibinfo{person}{Christine
  Bauer}, \bibinfo{person}{Wolfgang Reisinger}, \bibinfo{person}{Dominik
  Kowald}, {and} \bibinfo{person}{Elisabeth Lex}.}
  \bibinfo{year}{2020}\natexlab{}.
\newblock \bibinfo{booktitle}{\emph{{The dataset used in the article ``Listener
  Modeling and Context-aware Music Recommendation Based on Country
  Archetypes''}}}.
\newblock
\urldef\tempurl%
\url{https://doi.org/10.5281/zenodo.3907362}
\showDOI{\tempurl}


\bibitem[\protect\citeauthoryear{{Schedl} and {Ferwerda}}{{Schedl} and
  {Ferwerda}}{2017}]%
        {schedl_ism2017}
\bibfield{author}{\bibinfo{person}{Markus {Schedl}} {and}
  \bibinfo{person}{Bruce {Ferwerda}}.} \bibinfo{year}{2017}\natexlab{}.
\newblock \showarticletitle{Large-Scale Analysis of Group-Specific Music Genre
  Taste from Collaborative Tags}. In \bibinfo{booktitle}{\emph{2017 IEEE
  International Symposium on Multimedia}} (Taichung, Taiwan, 11--13 December)
  \emph{(\bibinfo{series}{ISM '17})}. \bibinfo{publisher}{IEEE},
  \bibinfo{pages}{479--482}.
\newblock
\urldef\tempurl%
\url{https://doi.org/10.1109/ISM.2017.95}
\showDOI{\tempurl}


\bibitem[\protect\citeauthoryear{Schedl, Knees, McFee, Bogdanov, and
  Kaminskas}{Schedl et~al\mbox{.}}{2015}]%
        {schedl_etal:rsh:2015}
\bibfield{author}{\bibinfo{person}{Markus Schedl}, \bibinfo{person}{Peter
  Knees}, \bibinfo{person}{Brian McFee}, \bibinfo{person}{Dmitry Bogdanov},
  {and} \bibinfo{person}{Marius Kaminskas}.} \bibinfo{year}{2015}\natexlab{}.
\newblock \showarticletitle{{Music Recommender Systems}}.
\newblock In \bibinfo{booktitle}{\emph{{Recommender Systems Handbook}}
  (\bibinfo{edition}{2nd} ed.)}, \bibfield{editor}{\bibinfo{person}{Francesco
  Ricci}, \bibinfo{person}{Lior Rokach}, \bibinfo{person}{Bracha Shapira},
  {and} \bibinfo{person}{Paul~B. Kantor}} (Eds.).
  \bibinfo{publisher}{Springer}, \bibinfo{pages}{453--492}.
\newblock


\bibitem[\protect\citeauthoryear{{Schedl}, {Lemmerich}, {Ferwerda}, {Skowron},
  and {Knees}}{{Schedl} et~al\mbox{.}}{2017}]%
        {schedl2017_ism}
\bibfield{author}{\bibinfo{person}{Markus {Schedl}}, \bibinfo{person}{Florian
  {Lemmerich}}, \bibinfo{person}{Bruce {Ferwerda}}, \bibinfo{person}{Marcin
  {Skowron}}, {and} \bibinfo{person}{Peter {Knees}}.}
  \bibinfo{year}{2017}\natexlab{}.
\newblock \showarticletitle{Indicators of Country Similarity in Terms of Music
  Taste, Cultural, and Socio-economic Factors}. In
  \bibinfo{booktitle}{\emph{2017 IEEE International Symposium on Multimedia
  (ISM)}} \emph{(\bibinfo{series}{ISM 2017})}. IEEE, \bibinfo{pages}{308--311}.
\newblock
\urldef\tempurl%
\url{https://doi.org/10.1109/ISM.2017.55}
\showDOI{\tempurl}


\bibitem[\protect\citeauthoryear{Schedl and Tkalcic}{Schedl and
  Tkalcic}{2014}]%
        {DBLP:conf/mm/SchedlT14}
\bibfield{author}{\bibinfo{person}{Markus Schedl} {and} \bibinfo{person}{Marko
  Tkalcic}.} \bibinfo{year}{2014}\natexlab{}.
\newblock \showarticletitle{{Genre-based Analysis of Social Media Data on Music
  Listening Behavior: Are Fans of Classical Music Really Averse to Social
  Media?}}. In \bibinfo{booktitle}{\emph{Proceedings of the First International
  Workshop on Internet-Scale Multimedia Management, {WISMM} '14, Orlando,
  Florida, USA, November 7, 2014}}, \bibfield{editor}{\bibinfo{person}{Roger
  Zimmermann} {and} \bibinfo{person}{Yi~Yu}} (Eds.).
  \bibinfo{publisher}{{ACM}}, \bibinfo{pages}{9--13}.
\newblock
\urldef\tempurl%
\url{https://doi.org/10.1145/2661714.2661717}
\showDOI{\tempurl}


\bibitem[\protect\citeauthoryear{Schedl, Zamani, Chen, Deldjoo, and
  Elahi}{Schedl et~al\mbox{.}}{2018}]%
        {Schedl2018}
\bibfield{author}{\bibinfo{person}{Markus Schedl}, \bibinfo{person}{Hamed
  Zamani}, \bibinfo{person}{Ching-Wei Chen}, \bibinfo{person}{Yashar Deldjoo},
  {and} \bibinfo{person}{Mehdi Elahi}.} \bibinfo{year}{2018}\natexlab{}.
\newblock \showarticletitle{Current challenges and visions in music recommender
  systems research}.
\newblock \bibinfo{journal}{\emph{International Journal of Multimedia
  Information Retrieval}} \bibinfo{volume}{7}, \bibinfo{number}{2}
  (\bibinfo{date}{June} \bibinfo{year}{2018}), \bibinfo{pages}{95--116}.
\newblock
\showISSN{2192-662X}
\urldef\tempurl%
\url{https://doi.org/10.1007/s13735-018-0154-2}
\showDOI{\tempurl}


\bibitem[\protect\citeauthoryear{Singhi and Brown}{Singhi and Brown}{2014}]%
        {Singhi2014_ismir}
\bibfield{author}{\bibinfo{person}{A. Singhi} {and} \bibinfo{person}{D.~G.
  Brown}.} \bibinfo{year}{2014}\natexlab{}.
\newblock \showarticletitle{On Cultural, Textual and Experiential Aspects of
  Music Mood}. In \bibinfo{booktitle}{\emph{Proceedings of the 15th
  International Society for Music Information Retrieval Conference}}
  \emph{(\bibinfo{series}{ISMIR '14})}. \bibinfo{address}{Taipei, Taiwan},
  \bibinfo{pages}{3--8}.
\newblock


\bibitem[\protect\citeauthoryear{Skowron, Lemmerich, Ferwerda, and
  Schedl}{Skowron et~al\mbox{.}}{2017}]%
        {Skowron2017_genreprediction}
\bibfield{author}{\bibinfo{person}{Marcin Skowron}, \bibinfo{person}{Florian
  Lemmerich}, \bibinfo{person}{Bruce Ferwerda}, {and} \bibinfo{person}{Markus
  Schedl}.} \bibinfo{year}{2017}\natexlab{}.
\newblock \showarticletitle{Predicting Genre Preferences from Cultural and
  Socio-Economic Factors for Music Retrieval}. In
  \bibinfo{booktitle}{\emph{Advances in Information Retrieval}},
  \bibfield{editor}{\bibinfo{person}{Joemon~M Jose}, \bibinfo{person}{Claudia
  Hauff}, \bibinfo{person}{Ismail~Sengor Alt{\i}ngovde}, \bibinfo{person}{Dawei
  Song}, \bibinfo{person}{Dyaa Albakour}, \bibinfo{person}{Stuart Watt}, {and}
  \bibinfo{person}{John Tait}} (Eds.). \bibinfo{publisher}{Springer
  International Publishing}, \bibinfo{address}{Cham, Germany},
  \bibinfo{pages}{561--567}.
\newblock
\showISBNx{978-3-319-56608-5}


\bibitem[\protect\citeauthoryear{Sonnett}{Sonnett}{2016}]%
        {SONNETT201638}
\bibfield{author}{\bibinfo{person}{John Sonnett}.}
  \bibinfo{year}{2016}\natexlab{}.
\newblock \showarticletitle{Ambivalence, indifference, distinction: A
  comparative netfield analysis of implicit musical boundaries}.
\newblock \bibinfo{journal}{\emph{Poetics}}  \bibinfo{volume}{54}
  (\bibinfo{year}{2016}), \bibinfo{pages}{38--53}.
\newblock
\showISSN{0304-422X}
\urldef\tempurl%
\url{https://doi.org/10.1016/j.poetic.2015.09.002}
\showDOI{\tempurl}


\bibitem[\protect\citeauthoryear{{Statista Research Department}}{{Statista
  Research Department}}{2019}]%
        {statista_uk}
\bibfield{author}{\bibinfo{person}{{Statista Research Department}}.}
  \bibinfo{year}{2019}\natexlab{}.
\newblock \showarticletitle{Music industry in the {U}nited {K}ingdom --
  Statistics \& Facts}.
\newblock  (\bibinfo{year}{2019}).
\newblock
\newblock
\shownote{{https://www.statista.com/topics/3152/music-industry-in-the-united-kingdom-uk/}.}


\bibitem[\protect\citeauthoryear{Stevens}{Stevens}{2012}]%
        {stevens2012}
\bibfield{author}{\bibinfo{person}{Catherine~J. Stevens}.}
  \bibinfo{year}{2012}\natexlab{}.
\newblock \showarticletitle{Music perception and cognition: a review of recent
  cross-cultural research}.
\newblock \bibinfo{journal}{\emph{Topics in Cognitive Science}}
  \bibinfo{volume}{4}, \bibinfo{number}{4} (\bibinfo{year}{2012}),
  \bibinfo{pages}{653--667}.
\newblock
\urldef\tempurl%
\url{https://doi.org/10.1111/j.1756-8765.2012.01215.x}
\showDOI{\tempurl}


\bibitem[\protect\citeauthoryear{{ter Bogt}, Delsing, van Zalk, Christenson,
  and Meeus}{{ter Bogt} et~al\mbox{.}}{2011}]%
        {terbogt2011}
\bibfield{author}{\bibinfo{person}{T.F.M. {ter Bogt}},
  \bibinfo{person}{M.J.M.H. Delsing}, \bibinfo{person}{M. van Zalk},
  \bibinfo{person}{P.G. Christenson}, {and} \bibinfo{person}{W.H.J. Meeus}.}
  \bibinfo{year}{2011}\natexlab{}.
\newblock \showarticletitle{Intergenerational continuity of taste: parental and
  adolescent music preferences}.
\newblock \bibinfo{journal}{\emph{Social Forces}} \bibinfo{volume}{90},
  \bibinfo{number}{1} (\bibinfo{year}{2011}), \bibinfo{pages}{297--319}.
\newblock


\bibitem[\protect\citeauthoryear{Tiwari, Pangtey, and Kumar}{Tiwari
  et~al\mbox{.}}{2018}]%
        {10.1007/978-3-319-95171-3_51}
\bibfield{author}{\bibinfo{person}{Sunita Tiwari},
  \bibinfo{person}{Manjeet~Singh Pangtey}, {and} \bibinfo{person}{Sushil
  Kumar}.} \bibinfo{year}{2018}\natexlab{}.
\newblock \showarticletitle{{Location Aware Personalized News Recommender
  System Based on Twitter Popularity}}. In
  \bibinfo{booktitle}{\emph{Computational Science and Its Applications -- ICCSA
  2018}}, \bibfield{editor}{\bibinfo{person}{Osvaldo Gervasi},
  \bibinfo{person}{Beniamino Murgante}, \bibinfo{person}{Sanjay Misra},
  \bibinfo{person}{Elena Stankova}, \bibinfo{person}{Carmelo~M. Torre},
  \bibinfo{person}{Ana Maria~A.C. Rocha}, \bibinfo{person}{David Taniar},
  \bibinfo{person}{Bernady~O. Apduhan}, \bibinfo{person}{Eufemia Tarantino},
  {and} \bibinfo{person}{Yeonseung Ryu}} (Eds.). \bibinfo{publisher}{Springer},
  \bibinfo{address}{Cham}, \bibinfo{pages}{650--658}.
\newblock
\showISBNx{978-3-319-95171-3}


\bibitem[\protect\citeauthoryear{Vall, Quadrana, Schedl, and Widmer}{Vall
  et~al\mbox{.}}{2019}]%
        {Vall2019}
\bibfield{author}{\bibinfo{person}{Andreu Vall}, \bibinfo{person}{Massimo
  Quadrana}, \bibinfo{person}{Markus Schedl}, {and} \bibinfo{person}{Gerhard
  Widmer}.} \bibinfo{year}{2019}\natexlab{}.
\newblock \showarticletitle{Order, context and popularity bias in next-song
  recommendations}.
\newblock \bibinfo{journal}{\emph{International Journal of Multimedia
  Information Retrieval}} \bibinfo{volume}{8}, \bibinfo{number}{2}
  (\bibinfo{date}{June} \bibinfo{year}{2019}), \bibinfo{pages}{101--113}.
\newblock
\showISSN{2192-662X}
\urldef\tempurl%
\url{https://doi.org/10.1007/s13735-019-00169-8}
\showDOI{\tempurl}


\bibitem[\protect\citeauthoryear{{van der Maaten} and Hinton}{{van der Maaten}
  and Hinton}{2008}]%
        {vandermaaten:jmlr:2008}
\bibfield{author}{\bibinfo{person}{Laurens {van der Maaten}} {and}
  \bibinfo{person}{Geoffrey Hinton}.} \bibinfo{year}{2008}\natexlab{}.
\newblock \showarticletitle{Visualizing Data using {t-SNE}}.
\newblock \bibinfo{journal}{\emph{Journal of Machine Learning Research}}
  \bibinfo{volume}{9} (\bibinfo{date}{November} \bibinfo{year}{2008}),
  \bibinfo{pages}{2579--2605}.
\newblock
\urldef\tempurl%
\url{http://www.jmlr.org/papers/volume9/vandermaaten08a/vandermaaten08a.pdf}
\showURL{%
\tempurl}


\bibitem[\protect\citeauthoryear{{van Venrooij}}{{van Venrooij}}{2009}]%
        {VANVENROOIJ2009315}
\bibfield{author}{\bibinfo{person}{Alex {van Venrooij}}.}
  \bibinfo{year}{2009}\natexlab{}.
\newblock \showarticletitle{The aesthetic discourse space of popular music:
  1985--86 and 2004--05}.
\newblock \bibinfo{journal}{\emph{Poetics}} \bibinfo{volume}{37},
  \bibinfo{number}{4} (\bibinfo{year}{2009}), \bibinfo{pages}{315--332}.
\newblock
\showISSN{0304-422X}
\urldef\tempurl%
\url{https://doi.org/10.1016/j.poetic.2009.06.005}
\showDOI{\tempurl}


\bibitem[\protect\citeauthoryear{Vigliensoni and Fujinaga}{Vigliensoni and
  Fujinaga}{2016}]%
        {Vigliensoni2016}
\bibfield{author}{\bibinfo{person}{G. Vigliensoni} {and} \bibinfo{person}{I.
  Fujinaga}.} \bibinfo{year}{2016}\natexlab{}.
\newblock \showarticletitle{Automatic music recommendation systems: do
  demographic, profiling, and contextual features improve their performance?}.
  In \bibinfo{booktitle}{\emph{17th International Society for Music Information
  Retrieval Conference}} \emph{(\bibinfo{series}{ISMIR '16})}.
  \bibinfo{pages}{94--100}.
\newblock


\bibitem[\protect\citeauthoryear{Vlegels and Lievens}{Vlegels and
  Lievens}{2017}]%
        {VLEGELS201776}
\bibfield{author}{\bibinfo{person}{Jef Vlegels} {and} \bibinfo{person}{John
  Lievens}.} \bibinfo{year}{2017}\natexlab{}.
\newblock \showarticletitle{Music classification, genres, and taste patterns: A
  ground-up network analysis on the clustering of artist preferences}.
\newblock \bibinfo{journal}{\emph{Poetics}}  \bibinfo{volume}{60}
  (\bibinfo{year}{2017}), \bibinfo{pages}{76--89}.
\newblock
\showISSN{0304-422X}
\urldef\tempurl%
\url{https://doi.org/10.1016/j.poetic.2016.08.004}
\showDOI{\tempurl}


\bibitem[\protect\citeauthoryear{Wilcoxon}{Wilcoxon}{1945}]%
        {wilcoxon}
\bibfield{author}{\bibinfo{person}{Frank Wilcoxon}.}
  \bibinfo{year}{1945}\natexlab{}.
\newblock \showarticletitle{Individual Comparisons by Ranking Methods}.
\newblock \bibinfo{journal}{\emph{Biometrics Bulletin}} \bibinfo{volume}{1},
  \bibinfo{number}{6} (\bibinfo{year}{1945}), \bibinfo{pages}{80--83}.
\newblock
\showISSN{00994987}
\urldef\tempurl%
\url{http://www.jstor.org/stable/3001968}
\showURL{%
\tempurl}


\bibitem[\protect\citeauthoryear{Zangerle, Pichl, and Schedl}{Zangerle
  et~al\mbox{.}}{2018}]%
        {zangerle_etal:umap:2018}
\bibfield{author}{\bibinfo{person}{Eva Zangerle}, \bibinfo{person}{Martin
  Pichl}, {and} \bibinfo{person}{Markus Schedl}.}
  \bibinfo{year}{2018}\natexlab{}.
\newblock \showarticletitle{{Culture-Aware Music Recommendation}}. In
  \bibinfo{booktitle}{\emph{{Proceedings of the 26th Conference on User
  Modeling, Adaptation and Personalization (UMAP 2018)}}}.
  \bibinfo{publisher}{ACM}, \bibinfo{address}{Singapore, Singapore},
  \bibinfo{pages}{357--358}.
\newblock
\urldef\tempurl%
\url{https://doi.org/10.1145/3209219.3209258}
\showDOI{\tempurl}


\end{thebibliography}

\end{document}